\documentclass[11pt,a4paper]{article}
\usepackage{jcappub}

\usepackage{amssymb,amsmath,amstext,amsfonts}
\usepackage{bbold}
\usepackage{bm}
\usepackage{graphicx}
\usepackage{xcolor}
\usepackage{overpic}
\usepackage{subcaption}

\newcommand{\beq}{\begin{equation}}
\newcommand{\eeq}{\end{equation}}
\newcommand{\bal}{\begin{aligned}}
\newcommand{\eal}{\end{aligned}}
\newcommand{\Lag}{{\mathcal{L}}}
\newcommand{\Mp}{M_{\rm Pl}}

\newcommand{\etaperp}{\eta_\perp}
\newcommand{\F}{Q_s}
\newcommand{\etamax}{\eta_\perp^{\textrm{max}}}
\newcommand{\Nf}{N_{\textrm{f}}}

\newcommand{\kf}{k_{\textrm{f}}}

\newcommand{\de}{\delta}

\newcommand{\dea}{1/2}
\newcommand{\deb}{1/4}
\newcommand{\deA}{0.5}
\newcommand{\deB}{0.25}
\newcommand{\etaA}{14}
\newcommand{\etaB}{28}
\newcommand{\deeta}{7}

\newcommand{\inr}{\textit{in}}
\newcommand{\outr}{\textit{out}}

\newcommand{\x}{\kappa}

\newcommand{\kstar}{k_\star}
\newcommand{\OmegaRatio}{\Delta}
\newcommand{\bk}{\boldsymbol{k}}

\newcommand{\s}{s}

\newcommand{\Sm}{S_{-}}
\newcommand{\Sp}{S_{+}}
\newcommand{\Spm}{S_{\pm}}
\newcommand{\X}{X}

\newcommand{\Ni}{N_{\textrm{i}}}
\newcommand{\ki}{k_{\textrm{i}}}

\newcommand{\Angle}{\Delta \theta}
\newcommand{\deltaeta}{\etaperp \delta}

\newcommand{\af}{a_{\textrm{f}}}

\newcommand{\rhob}{\rho_{\textrm{b}}}

\newcommand{\Mn}{\Omega}

\newcommand{\Pzero}{\mathcal{P}_{0}}
\newcommand{\PCMB}{{\cal P}_{{\textsc{cmb}}}}

\newcommand{\omegalin}{\omega_{\textrm{lin}}}
\newcommand{\omegagwlin}{\omega_{\textrm{lin}}^{\textsc{gw}}}
\newcommand{\omegalog}{\omega_{\textrm{log}}}
\newcommand{\omegagwlog}{\omega_{\textrm{log}}^{\textsc{gw}}}
\newcommand{\omegalogc}{\omega_{\textrm{log,c}}}

\newcommand{\Omegainf}{\Omega_{\textrm{GW}}^{{\mathrm{Inf}}}}
\newcommand{\Omegarad}{\Omega_{\textrm{GW}}^{\mathrm{rad}}}

\title{Oscillations in the stochastic gravitational wave background from sharp features and particle production during inflation}

\author{Jacopo Fumagalli,}
\author{S\'{e}bastien Renaux-Petel}
\author{and Lukas~T.~Witkowski}

\affiliation{Institut d'Astrophysique de Paris, GReCO, UMR 7095 du CNRS et de Sorbonne Universit\'{e},\\ 98bis
boulevard Arago, 75014 Paris, France}

\emailAdd{jacopo.fumagalli@iap.fr}
\emailAdd{renaux@iap.fr}
\emailAdd{lukas.witkowski@iap.fr}

\abstract{We identify a characteristic pattern in the scalar-induced stochastic gravitational wave background from particle production during inflation.
If particle production is sufficiently efficient, the scalar power spectrum exhibits $\mathcal{O}(1)$ oscillations periodic in $k$, characteristic of a sharp feature, with an exponentially enhanced envelope. We systematically study the properties of the induced spectrum of gravitational waves sourced after inflation and find that this inherits the periodic structure in $k$, resulting in a peak in the gravitational wave energy density spectrum with $\mathcal{O}(10 \%)$ modulations. The frequency of the oscillation in the scalar power spectrum is determined by the scale of the feature during inflation and in turn sets the frequency of modulations in the gravitational wave signal. We present an explicit realisation of this phenomenon in the framework of multifield inflation, in the form of a strong sharp turn in the inflationary trajectory. The resulting stochastic background is potentially detectable in future gravitational wave observatories, and considerations of backreaction and perturbativity can be used to constrain the parameter space from the theoretical side. Our work motivates more extensive research linking primordial features to observable properties of the stochastic background of gravitational waves, and dedicated development in data analysis for their detection.}

\begin{document}

\maketitle

\section{Introduction}
\label{sec:intro}

A primary objective of the nascent field of gravitational wave (GW) astronomy is the detection of a stochastic gravitational wave background (SGWB). The SGWB is expected to receive contributions from astrophysical phenomena (reviewed in \cite{Regimbau:2011rp}), but also from cosmological processes in the early universe, making it an important probe of early universe physics alongside the cosmic microwave background (CMB), large-scale structure (LSS) and primordial black holes (PBHs).

Cosmological contributions to the SGWB arise from phase transitions, topological defects, tensor fluctuations sourced during inflation or preheating, and GWs induced by primordial scalar fluctuations in the post-inflationary era, see \cite{Caprini:2018mtu, Mazumdar:2018dfl} for recent reviews of these topics. The latter originate from the fact that primordial density perturbations act as a source for gravitational waves when they reenter the Hubble radius during the radiation era
\cite{Ananda:2006af, Baumann:2007zm}.

At scales $\gtrsim 1$ Mpc the scalar fluctuations are already tightly constrained by CMB measurements and LSS to be approximately Gaussian with a near scale-invariant power spectrum $\mathcal{P}_\zeta \sim 10^{-9}$, resulting in scalar-induced GWs with an amplitude that is too small for detection. 

Much less is known at short scales ($\ll 1$ Mpc) where the CMB constraints do not apply. This explicitly permits scenarios where the scalar fluctuations associated to these modes are significantly enhanced compared to the CMB ones, leading to potentially detectable GWs.
These scalar-induced GWs will contain information about the scalar fluctuations and in particular the scalar power spectrum $\mathcal{P}_\zeta$, as the energy density power spectrum of scalar-induced GWs is related to the scalar power spectrum as $\Omega_{\textrm{GW}} \sim \iint \mathcal{P}_\zeta^2$ \cite{Ananda:2006af, Baumann:2007zm}, see also \cite{Acquaviva:2002ud, Mollerach:2003nq} for related earlier work. This leads to the exciting prospect that measurements of the SGWB can give direct information about the primordial scalar fluctuations at small scales.

\begin{figure}[t]
\centering
\begin{overpic}[width=1.0\textwidth]{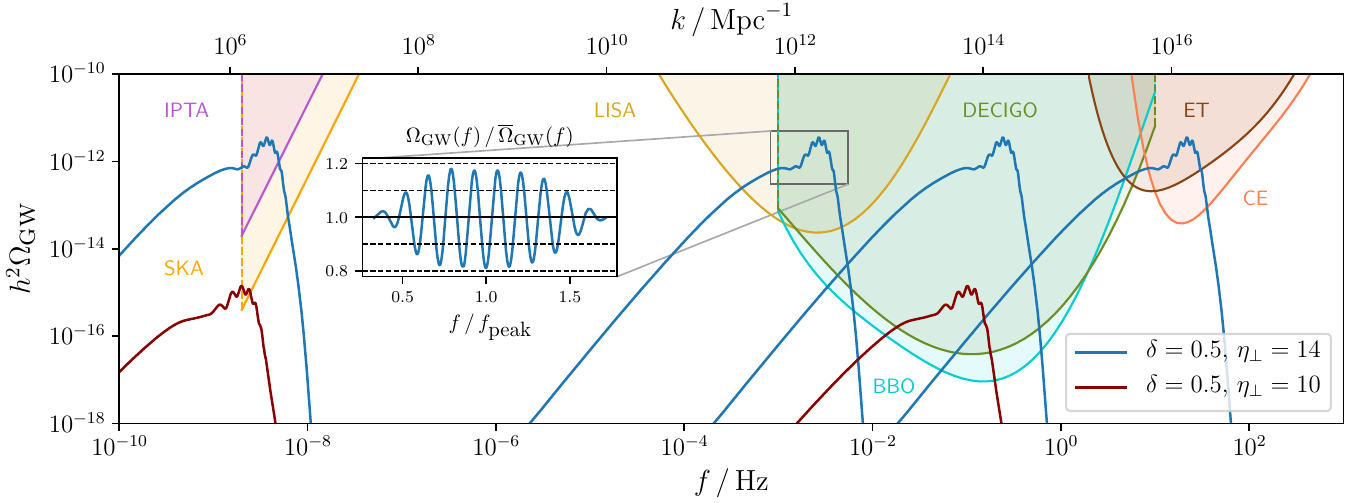}
\end{overpic}
\caption{\textit{Distinctive stochastic GW profile from sharp features leading to copious particle production, plotted together with power-law-integrated sensitivity curves for a
selected choice of GW observatories taken from  \cite{Schmitz:2020syl}. A key property of the GW energy density power spectrum $h^2 \Omega_{\textrm{GW}}$ 
is a $\mathcal{O}(10 \%)$-level periodic modulation on the principal peak, as can be seen more clearly in the inset, where we plot the ratio  $\Omega_{\textrm{GW}} (f) / \overline{\Omega}_{\textrm{GW}} (f)$ with $\overline{\Omega}_{\textrm{GW}}(f)$ the GW signal with the modulations smoothed out.  The choice of parameters in the legend are for multifield inflationary scenarios where the sharp feature is due to a strong turn of the inflationary trajectory. The mechanism and its parameterisation will be discussed in detail in sec.~\ref{sec:setup}. For reference, the model parameters for the plots shown are $(\delta, \etaperp)=(\deA,\etaA)$ with $N_\textrm{f}=15, \, 28.5, \, 33, \, 37$ (blue curves), and $(\delta, \etaperp)=(0.5,10)$ with $N_\textrm{f}=14.75, \, 32.5$ (dark red curves).}}
\label{fig:Omega-of-f-sensitivity}
\end{figure}

The enhancement of scalar fluctuations required for detectable GWs necessitates a departure of inflation from the standard single-field slow-roll paradigm, making the SGWB a key target for testing more complicated implementations of inflation, as arise in high energy theory embeddings. This top-down perspective motivates inflationary mechanisms where multiple fields can become dynamical during inflation or where inflation may occur in several distinct phases (see e.g.~the recent work \cite{DAmico:2020euu}). Interestingly, departures from single-field slow-roll are generically associated with so-called \emph{features} (see the reviews \cite{Chen:2010xka,Chluba:2015bqa,Slosar:2019gvt} and references therein), which in this context refers to oscillatory patterns in the correlation functions of the primordial density fluctuations. The expectation thus is that inflation models with a peak in $\mathcal{P}_\zeta$, as required for producing detectable GWs, will also naturally exhibit a feature, i.e.~an oscillatory modulation of that peak. 

This motivates a research programme dedicated to studying features in $\mathcal{P}_\zeta$ at small scales through their imprint on $\Omega_{\textrm{GW}}$. Features of the primordial curvature power spectrum represent a longstanding theoretical interest due to their relation to particular models of inflation, their role for testing the inflationary paradigm against alternatives and as signatures for the existence of heavy particles beyond the reach of terrestrial experiments (see again the reviews \cite{Chen:2010xka,Chluba:2015bqa,Slosar:2019gvt})
A key difference from studies of features at CMB scales is that, to lead to detectable GWs, the scalar fluctuations at the scale of the feature also need to be amplified. At the same time, in the absence of strong constraints, a richer kaleidoscope of features can be envisaged. 

Here we take first steps in this direction by focussing on what is referred to as \emph{sharp features}, i.e. $\mathcal{P}_\zeta (k)$ with oscillations that are periodic in $k$, signaling a localised event of particle production. As we will review, if particles are produced with sufficiently large occupation numbers, the power spectrum $\mathcal{P}_\zeta$ will be amplified accordingly, together with the amplitude of oscillations. 
In this paper, we will concentrate on a mechanism leading to copious particle production ($n_k \gg1$) ensuring that the boost of the power spectrum is sufficient to lead to a detectable SGWB,
and which is automatically accompanied by $\mathcal{O}(1)$ oscillations in $\mathcal{P}_\zeta (k)$. Such $\mathcal{O}(1)$ modulations can be realised with more modest occupation numbers $n_k \sim 1$, but the required enhancement of scalar fluctuations must then come from a different mechanism.

One of the main results of this work is the description of the characteristics of the scalar-induced GW spectrum $\Omega_\textrm{GW}(k)$ that arises from a sharp feature. We observe that the overall shape of $\Omega_\textrm{GW}(k)$ is governed by the envelope of the scalar power spectrum and generically consists of a taller principal peak and potentially a lower broad peak for smaller values of $k$. Most importantly, the principal peak in $\Omega_\textrm{GW}(k)$ exhibits periodic modulations, with a frequency larger by a factor $\sqrt{3}$ than that of $\mathcal{P}_\zeta(k)$.
The $\mathcal{O}(1)$ oscillations of $\mathcal{P}_\zeta(k)$ are translated into a $\mathcal{O}(10 \%)$ modulation in $\Omega_\textrm{GW}(k)$, which can be understood as an `averaging-out' effect tied to the physics of how scalar fluctuations are processed into tensor modes. This can be seen explicitly in fig.~\ref{fig:Omega-of-f-sensitivity}, where the blue and dark red curves denote $\Omega_\textrm{GW}$ computed for an explicit realisation of a sharp feature (plotted against frequency $f$ which is linearly related to $k$.)

We find that in the vicinity of the principal peak the GW energy density power spectrum can be modelled by the template
\beq
\Omega_\textrm{GW}(k)=\overline{\Omega}_{\textrm{GW}}(k) \left(1+A \cos( \omegagwlin k+\phi)  \right) \, ,
\eeq
i.e.~it is given by a sinusoidal modulation with amplitude $A \sim \mathcal{O}(10 \%)$ and frequency $\omegagwlin$ about a smooth background $\overline{\Omega}_{\textrm{GW}}$, which corresponds to the GW signal with the modulations averaged out. This can be seen in the inset in fig.~\ref{fig:Omega-of-f-sensitivity}) where we plot the ratio $\Omega_\textrm{GW} / \overline{\Omega}_{\textrm{GW}}$ as a function of $f/f_{\textrm{peak}} = k/k_{\textrm{peak}}$ for the model depicted by the blue curve. Depending on the exact realisation of the sharp feature, the precise form of the background $\overline{\Omega}_{\textrm{GW}}(k)$ will vary (e.g.~compare the blue and dark red curves in fig.~\ref{fig:Omega-of-f-sensitivity}), but the properties of the modulations are robust. Close to the peak both $A$ and $\omegagwlin$ are near-constant, but especially $A$ and also $\omegagwlin$ become $k$-dependent away from the centre of the principal peak.

As an explicit realisation of a sharp feature we will consider  multifield inflationary scenarios with a strong sharp turn in the inflationary trajectory, i.e.~with a strong departure of the trajectory from a geodesic for at most one $e$-fold. Sharp turns in inflation have been extensively studied in the past years, see e.g.~\cite{Achucarro:2010da,Achucarro:2010jv,Chen:2011zf,Shiu:2011qw,Cespedes:2012hu,Achucarro:2012sm,Avgoustidis:2012yc,Gao:2012uq,Achucarro:2012fd,Konieczka:2014zja}, albeit not in contexts giving rise to copious particle production and large enhancement of the power spectrum that are central for this paper. Strong sharp turns as a means of enhancing the scalar power spectrum at small scales have only been considered recently in \cite{Palma:2020ejf,Fumagalli:2020adf} in the context of primordial black hole production. Here we compute the scalar power spectrum both numerically and analytically for various realisations of a strong sharp turn, finding that the scalar power spectrum in all cases is characterised by periodic oscillations in $k$, typical of a sharp feature, with $\mathcal{O}(1)$ amplitude of oscillations and a localised exponentially enhanced envelope.
Boosted and oscillatory power spectra have also been observed to arise in single-field models in \cite{Ballesteros:2018wlw,Kefala:2020xsx}, and we expect our results to be broadly applicable there as well. Strong broad turns also lead to an amplification of scalar fluctuations \cite{Braglia:2020eai,Fumagalli:2020adf, Aldabergenov:2020bpt}, but the power spectrum does not exhibit oscillations and hence they do not realise a sharp feature of interest here.

We find that the scalar-induced contributions to the SGWB from strong sharp turns during inflation can be potentially observable by various future GW observatories for different choices of model parameters, see once again fig.~\ref{fig:Omega-of-f-sensitivity}. The enhancement of scalar fluctuations and hence the maximal amplitude of $h^2 \Omega_\textrm{GW}$ is generically bounded in these models if excessive backreaction on the inflationary background and a loss of perturbative control are to be avoided. A signal with $h^2 \Omega_\textrm{GW} \sim 10^{-13}$, as needed for detection by the space-based interferometer LISA, will generically be safe from backreaction, but may require a more detailed assessment regarding perturbative control than the tentative analysis performed here. GW spectra with $h^2 \Omega_\textrm{GW} \sim 10^{-15}$, that are expected to be visible in DECIGO and BBO, are found to be consistent with backreaction and perturbativity constraints. The regime of theoretical control can be further expanded if the `base' value of the power spectrum, i.e.~the power spectrum in the absence of the strong turn, is boosted at small scales compared to its value at CMB scales independently of the turn. Eventually, we show that for theoretically viable models, the tensor perturbations generated during inflation by the burst of particle production lead to a subdominant contribution to the SGWB compared the one processed in the radiation era.

We envisage several directions for future work. One concerns the continuation of the programme of studying features in $\mathcal{P}_\zeta$ through their GW signatures. Besides sharp features considered here, two other frequently studied types are \emph{resonant features} and \emph{primordial standard clocks}, and it would be interesting to investigate their imprint on the SGWB. Preliminary results for resonant features reported here show that they also lead to characteristic features in the SGWB, with GW spectra that exhibit structures periodic in $\log k$: 
\begin{align}
\label{eq:resonant_template_intro}
\Omega_{\textrm{GW}}(k) = \overline{\Omega}_{\textrm{GW}}(k) \Big[1 &+ A_{1} \cos \big(\omega_\textrm{log} \log (k/k_\textrm{ref}) + \phi_{1} \big) \\
\nonumber & + A_{2} \cos \big(2 \omega_\textrm{log} \log (k/k_\textrm{ref}) + \phi_{2} \big) \Big] \, ,
\end{align}
where $\omega_\textrm{log}$ is the frequency of oscillations of the scalar power spectrum, $k_\textrm{ref}$ is an arbitrary reference scale and $\overline{\Omega}_{\textrm{GW}}$ again denotes a smooth background. A more detailed analysis of this case will be the subject of a dedicated publication \cite{resonant-SGWB}. A complementary line of research is to investigate embeddings of sharp turns in microscopic models of inflation. A generic turn during inflation is often accompanied by a violation of the slow-roll conditions which can have additional effects on $\mathcal{P}_\zeta$ that are not of the form of a sharp feature.

A key result of our work is that, independently of the precise realisation studied here, features in the frequency profile of SGWBs are an interesting theoretically motivated target. We find it hence important to assess their detectability with LISA and other future GW observatories, and more generally to develop dedicated data analysis techniques for probing features with stochastic gravitational wave backgrounds.

The paper is structured as follows. To better illustrate our results we will frequently employ an explicit realisation of a sharp feature in the framework of inflation with a strong and sharp turn in field space. Hence, Sec.~\ref{sec:model} is dedicated to the presentation of this framework, its setup and dynamics (Sec.~\ref{sec:setup}), the enhancement of fluctuations due to a strong and sharp turn (Sec.~\ref{sec:enhancingP}), and constraints on this enhancement from backreaction effects and perturbativity considerations (Sec.~\ref{sec:control}). The observed peak in the power spectrum with $\mathcal{O}(1)$ modulations is not specific to this model, but a generic property of a sharp feature associated with copious particle production, as we explain in Sec.~\ref{sec:generic}. Sec.~\ref{sec:stochGWs} is dedicated to the characteristics of the GW energy density %fraction 
spectrum due to a sharp feature.
In Sec.~\ref{sec:GWproperties} we describe the resulting spectral shape of $\Omega_{\textrm{GW}}$, with findings that are applicable to any primordial scalar power spectrum that displays a sharp feature, while in Sec.~\ref{sec:GWpredictions} we consider its observational prospects. Sec.~\ref{sec:discussion} contains a summary of our results and an extensive outlook on future research directions. Some technical results are collected in appendix \ref{sec:analytics}.

\section{Enhanced oscillatory scalar power spectrum from sharp features and copious particle production}
\label{sec:model}

As explained in the introduction, the oscillating feature in the frequency profile of the GW energy density power spectrum that we highlight in this paper is a direct consequence of oscillations in the primordial power spectrum of the curvature perturbation. The latter are expected in many different scenarios in the early universe, for instance features in the inflationary potential/Lagrangian or sharp turns of the inflationary trajectory, and can be seen as the consequence of particle production/excited initial states for the relevant wavenumbers (see, e.g., Ref.~\cite{Chen:2010xka,Chluba:2015bqa,Slosar:2019gvt} for reviews on the subject). We thus expects our results to have a broad range of applicability. For definiteness, and as a first step in the direction of probing with gravitational waves copious particle production from features on small scales, we consider the setup recently studied in Refs.~\cite{Palma:2020ejf,Fumagalli:2020adf} of a sharp turn in two-field models of inflation leading to a large enhancement of the scalar power spectrum. In \ref{sec:setup} we present this setup, defining our notations and spelling out our assumptions. In \ref{sec:enhancingP}, building on the analytical result of \cite{Palma:2020ejf} and generalising it, we characterise the enhancement and the oscillations of the power spectrum by giving a simple analytical formula describing it, and we explain its characteristic patterns. We argue in \ref{sec:generic} that the latter are generic to models with a boosted power spectrum coming from a sharp feature. In \ref{sec:control} we qualitatively discuss the questions of backreaction and perturbativity in these setups.

\subsection{Inflation with strong turns in field space}
\label{sec:setup}

{\bf Setup.---} We consider the general class of nonlinear sigma models, typical of embeddings of inflation in high-energy physics, whose action reads
\beq \label{eq:action1}
S=\int d^4x\sqrt{-g}\bigg[\frac{\Mp^2}{2}\,R-\frac{1}{2}\,G_{IJ}\nabla^{\mu}\phi^I\nabla_{\mu}\phi^J-V(\phi)\bigg]\,,
\eeq
where the first term is the Einstein-Hilbert action, and $G_{IJ}(\phi)$ defines a metric in the internal field space parametrised by the coordinates $\phi^I$. We consider two fields for simplicity, making use of the convenient adiabatic-entropic basis in field space, defined along the background trajectory by $e^I_{\sigma}\equiv \dot{\phi}^I/\dot{\sigma}$ and $e^I_s$, which is orthogonal to $e^I_{\sigma}$, and with a definite orientation for the basis $(e^I_{\sigma},e^I_s)$. Here, $\dot{\sigma} \equiv (G_{IJ}\dot{\phi}^I\dot{\phi}^J)^{1/2}$ is related to $\epsilon \equiv -\dot{H}/H^2$ by $\epsilon=\dot \sigma^2/(2 H^2 \Mp^2)$, where $H(t)$ denotes the Hubble scale. The degree of ``bending'' of the trajectory is characterised by the dimensionless parameter $\etaperp$ such that
\beq
{\cal D}_t e^I_{\sigma}=H\eta_{\perp}e^I_s\,,
\eeq
where the time field space covariant derivative of any field space vector $A^I$ is defined as ${\cal D}_t A^I=\dot{A}^I+\Gamma^I_{JK}\dot{\phi}^JA^K$. It indeed measures the amount of deviation of the background trajectory from a field space geodesic \cite{GrootNibbelink:2000vx,GrootNibbelink:2001qt}, or equivalently the acceleration of the trajectory perpendicular to its velocity. It is naturally associated to the ``turn rate'' of the trajectory, the energy scale $\Mn=H \etaperp$, such that a strong turn with $\etaperp \gg 1$, of interest in this paper, corresponds to a typical energy scale $\Mn \gg H$.

The physics of scalar linear fluctuations about the above background is described by the following quadratic action \cite{Sasaki:1995aw,GrootNibbelink:2001qt,Langlois:2008mn} (writing $S=\int {\rm d}t \,{\rm d}^3 x {\cal L}$)
\begin{eqnarray}
\Lag^{(2)}=a^3\bigg[\Mp^2\epsilon\left(\dot{\zeta}^2-\frac{(\partial \zeta)^2}{a^2}\, \right)+2\dot{\sigma}\eta_{\perp}\dot{\zeta}\F +\frac{1}{2}\left(\dot{\F}^2-\frac{(\partial \F)^2}{a^2}-m_s^2\F^2\right)\bigg]\,.
\label{L2}
\end{eqnarray}  
Here, $\zeta$ is the comoving curvature perturbation of observational interest, and $\F$ is the instantaneous entropic fluctuation, corresponding to fluctuations away from the background trajectory. In the comoving gauge, the field fluctuations read $\delta \phi^I=\F e_s^I$, while the spatial part of the metric takes the standard form $g_{ij}=a^2 e^{2 \zeta} \delta_{ij}$. The entropic mass term reads
\beq
\label{ms2}
m_s^2=V_{;ss}-H^2\eta_{\perp}^2+\epsilon H^2  \Mp^2 R_{\rm fs},
\eeq
with $V_{;ss}=e_s^I e_s^J V_{;IJ}$ the projection of the covariant Hessian of the potential along the entropic direction, and $R_{{\rm fs}}$ the field space scalar curvature. The Lagrangian \eqref{L2} is exact, and using the number of \textit{e}-folds of inflation $N=\ln(a)$ as the time variable, it shows that the physics of linear fluctuations in any two-field model is described by three functions of time: the Hubble scale $H(N)$, the entropic mass $m_s^2(N)$, and the bending $\etaperp(N)$. Based on this, Refs.~\cite{Palma:2020ejf,Fumagalli:2020adf} recently studied in a model-independent manner the enhancement of the curvature power spectrum present in models with a transient large bump in $\etaperp(N)$, as motivated by the ubiquitous phenomenon of turns in the inflationary landscape.
As can be seen from the expression \eqref{ms2}, a large bending can naturally lead to a negative entropic mass squared. As we discuss below, this does not imply a background instability, but rather a transient exponential growth of fluctuations, which will be central to this paper.
For simplicity, the Hubble scale $H(N)$ is assumed to be smooth and slowly-varying. We will come back later to this assumption. Without multifield effects, the curvature power spectrum would take the standard form
\beq
\mathcal{P}_{0}(k)=\frac{H^2}{8 \pi^2 \epsilon \Mp^2}\,,
\label{eq:P0}
\eeq
where all quantities on the right are evaluated at Hubble crossing $k=aH$, and hence with a mild scale dependence that we can neglect for our purposes. The enhancement of the power spectrum caused by the turn will be measured with respect to the single-field like power spectrum \eqref{eq:P0}.\\

\noindent {\bf Turn and characteristic scales.---} 
The turns in the inflationary trajectory that we will consider can be essentially characterised by three parameters: the time $\Nf$ of the feature, for definiteness measured after Hubble exit of the CMB pivot scale; its duration in \textit{e}-folds $\de$; and the typical value of $\etaperp$ during the turn, say the maximum value $\etamax$. In this paper, we will mostly consider a simple top-hat time-dependence of the bending parameter:
\begin{equation}
\eta_{\perp}(N)=\etamax \bigg[\theta\left(N-\left(\Nf-\frac{\delta}{2}\right)\right)-\theta\left(N-\left(\Nf+\frac{\delta}{2}\right)\right)\bigg]\,,\label{etaperp-top-hat}
\end{equation}
together with a temporarily negative entropic mass squared during the turn:
\beq
\frac{m_s^2}{H^2}=(\xi-1) \etaperp^2(N)  \qquad \textrm{during the turn}\,,
\label{eq:ms2-turn}
\eeq
with $\xi<1$ a constant parameter discussed below in more detail. As we will explain below from generic arguments, and as confirmed by explicit numerical computations, for the sharp turns of interest in this paper, the main features of the curvature power spectrum, and hence of the resulting spectrum of induced GWs, are insensitive to details of the time-dependence of $\etaperp(N)$ during the turn, as long as $\etaperp(N)$ can be characterised by a typical amplitude. 
Concrete models of turns are expected to display a smooth time-dependence of $\etaperp(N)$, like e.g.~a Gaussian time-profile
\begin{equation}
\eta_{\perp}(N)=\etamax e^{-(N-\Nf)^{2}/(2\de^{2})}\,,\label{etaperp-Gaussian}
\end{equation}
which we will use for comparison.  However, for our purposes, no artefact is introduced by considering the simple top-hat time-dependence \eqref{etaperp-top-hat}, which has the advantage of being amenable to simple analytical manipulations.\footnote{Using the WKB approximation, the computations performed for the top-hat model can also be generalized to any slowly varying time-dependence. Note also that the discontinuity of the top-hat model does introduce artificial oscillations up to arbitrarily large $k$, but they are suppressed and play no role in the following analysis. The oscillations we will discuss for the enhanced scales are physical, and are also present for smooth time-dependences like the one in Eq.~\eqref{etaperp-Gaussian}.} Hence, except otherwise stated, we focus on the latter in the following, and we will simply denote by $\etaperp$ the constant value of the bending parameter during the turn.\\

A turn in field space, as any ``feature'', has vastly different consequences depending on the scales of interest.
\begin{itemize}
    \item The scales most affected are the ones of energy $k/\af \sim \Mn$ around the time of the feature, hence their wavenumbers are of order $k \sim \kstar =\kf\, \etaperp$, where $\kf=\af H$ denotes the scale that crosses the Hubble radius at the (central) time of the feature $\Nf$. These scales are hence deep inside the Hubble radius when they ``hit'' the feature, and the corresponding curvature perturbation solely depends on $\etaperp^2$ and $m_s^2$ during the turn.
    
    \item Smaller scales $k \gg \kstar$ are even further inside the Hubble radius at $\Nf$, and hence are not affected by the turn, resulting in ${\cal P}_\zeta(k)=\Pzero$. 
    \item   Much larger scales $k \ll \kstar$ (typically) have already exited the Hubble radius at $\Nf$, and the precise impact that the turn has on them depends on the details of the past behaviour, but as we will see, it is much less pronounced than the impact on scales $k \sim \kstar$.
\end{itemize}

\begin{figure}[t]
\centering
\begin{subfigure}{.5\textwidth}
 \centering
   \begin{overpic}
[width=1.0\textwidth]{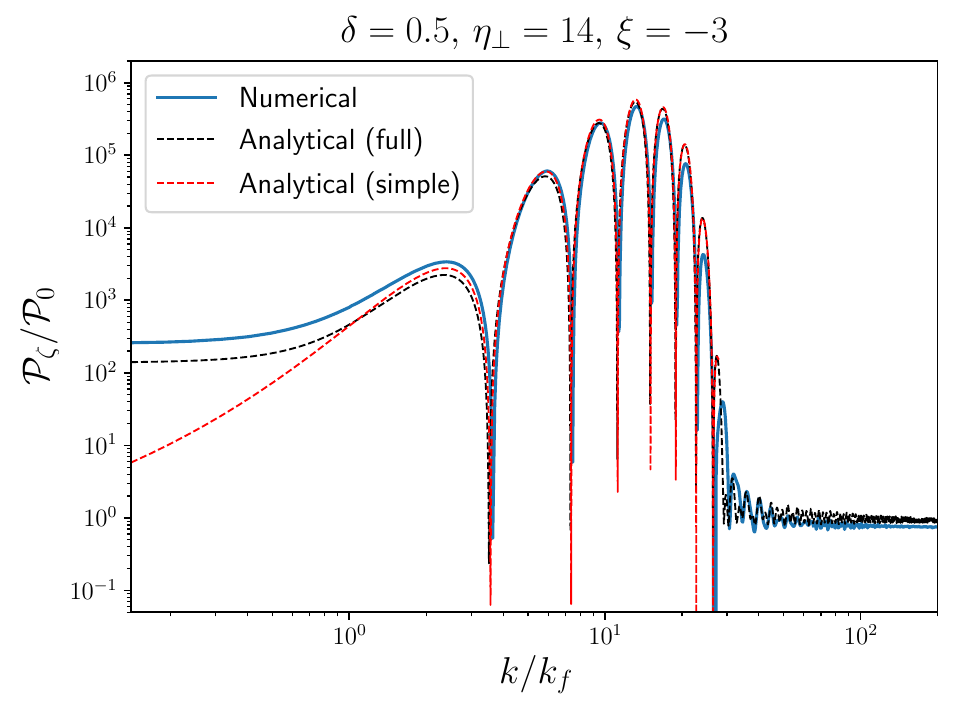}
\end{overpic}
\caption{\hphantom{A}}
\label{fig:figure1a}
\end{subfigure}%
\begin{subfigure}{.5\textwidth}
 \centering
   \begin{overpic}
[width=1.0\textwidth]{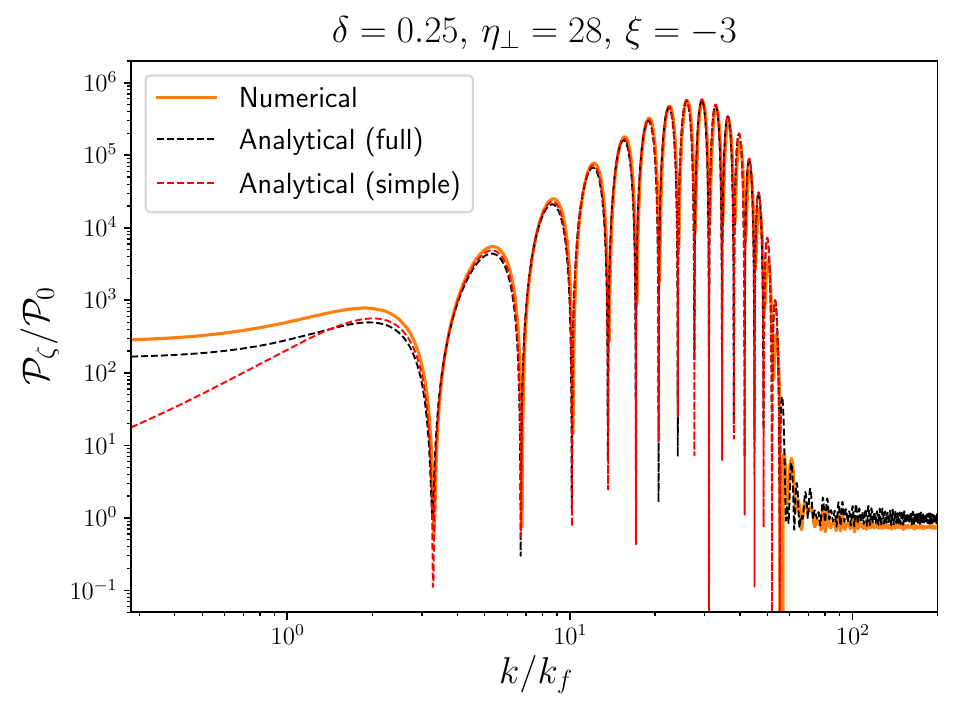}
\end{overpic}
\caption{\hphantom{A}}
\label{fig:figure1b}
\end{subfigure}%
\caption{
\textit{The curvature power spectrum generated by strong sharp turns discussed in the main text, with parameters $(\delta=0.5,\etaperp=14)$ (left) and $(\delta=0.25,\etaperp=28)$ (right), $\xi=-3$, and $m_s^2$ assumed to vanish before the turn. Plain curves correspond to numerical results. Dashed curves denote full and simplified analytical results. Oscillations in $k$ of the power spectrum are characteristics of sharp features, here with a localised exponentially enhanced envelope. Note that at large scales $\mathcal{P}_\zeta$ is enhanced compared to $\mathcal{P}_0$ (see the numerical and full analytical results), as a result of the choice $m_s^2=0$ before the turn. For the more generic situation of a non-vanishing entropic mass before the turn, the main peak is unaffected, but the enhancement at large scales is not present and instead one finds $\mathcal{P}_\zeta (k \ll k_\textrm{f}) \simeq \mathcal{P}_0$. See the main text for details.}}
\label{fig:examples}
\end{figure}

These general considerations are well illustrated by the result of numerical computations of the power spectrum in typical examples of interest for this paper, shown in fig.~\ref{fig:examples}. There, two set of parameters $(\delta=0.5,\etaperp=14)$ and $(\delta=0.25,\etaperp=28)$ have been chosen, together with $\xi=-3$, and $m_s^2$ was assumed to vanish before the turn. Plain curves correspond to numerical results, while black dashed curves denote analytical results found in \cite{Palma:2020ejf}, discussed in detail and generalised below, and red dashed curves denote a simplified analytical formula that is valid for the main scales of interest $k \sim \kstar$. The power spectrum exhibits characteristic oscillatory patterns and the scales $k \sim \kstar$ most affected by the turn are exponentially amplified compared to the baseline result $\Pzero$, while smaller scales are amplified, but to a much lesser extent. Once we come to observational consequences and theoretical constraints, the value of $\Pzero$ will matter, and as fig.~\ref{fig:examples} shows, its value is (indirectly) related to the value taken by the power spectrum on scales $k \ll \kstar$, and hence to the physics before the turn, 
which we now discuss.\\

\noindent {\bf Pre-turn physics and normalisation.---} Let us recall that on general grounds, one deduces from the action \eqref{L2} that on super-Hubble scales the entropic fluctuation evolves on its own, obeying $\ddot{\F}+ 3H \dot{\F}+m_{s,({\rm eff})}^{2} \F \simeq 0$, with an effective super-Hubble mass $m_{s({\rm eff})}^2=m_{s}^{2}+4\eta^2_{\perp}(N) H^{2}$ to which the bending contributes positively, and feeding the curvature perturbation, with $\dot{\zeta}\simeq-2H^2\eta_\perp/\dot{\sigma}\,\F$. In our paper \cite{Fumagalli:2020adf}, we considered that the Hessian and geometrical contribution to $m_s^2$ in Eq.~\eqref{ms2} are approximately constant throughout the evolution, so that $m_s^2/H^2=b-\etaperp^2(N)$, with $b={\cal O}(1)$ or larger, corresponding to the typical situation of a field with a non-negligible mass, before and after the turn. Then, the entropic fluctuations, for scales that have exited the Hubble radius before the feature, have decayed by that time, and decay even more during the turn, leaving no imprint on the large scale curvature power spectrum: ${\cal P}_\zeta \simeq \Pzero$ for $k \ll \kf$. 
On the other hand, Ref.~\cite{Palma:2020ejf} considers setups in which the various contributions to $m_s^2$ are such that $m_s^2/H^2=-4 \etaperp^2(N)$ at all times, and hence $m_{s({\rm eff})}^2=0$. In this so-called ultralight situation \cite{Achucarro:2016fby}, $Q_s \simeq H/(2 \pi)$ is constant on super-Hubble scales, giving ${\cal P}_\zeta \simeq \Pzero \left(1+4 \Angle^2 \right)$ for $k \ll \kf$, where $\Angle=\int \etaperp(N) d N$ denotes the total ``angle'' swept by the turning trajectory. With the entropic field being exactly massless before the turn, the analytical understanding is simplified in this situation. This is what has been assumed in fig.~\ref{fig:examples} leading to the enhancement at large scales there.

Yet another question concerns the relationship between the effective description around the time of the turn, which can arise tens of \textit{e}-folds after CMB scales exit the Hubble radius, and the physics active then. The phase of inflation may well have very different characteristics at these different epochs, with different values of $\epsilon$ for instance.
Hence, even if ${\cal P}_\zeta \simeq \Pzero$ for $k \ll \kf$, this description of the power spectrum for such large scales \textit{compared to the ones affected be the turn}, need not apply for the yet even larger scales probed by the CMB, at which we measure $\PCMB=2.4 \cdot 10^{-9}$. As a result, in the following, we may consider $\Pzero$ as an essentially free parameter, although for definiteness, the simple normalisation $\mathcal{P}_0=\PCMB$ is assumed in the plots if not otherwise stated.\\

\noindent {\bf Mass term.---} Let us now explain the rationale behind the parameterisation \eqref{eq:ms2-turn}. There, $\xi$ quantifies the relative importance during the turn of the Hessian and geometrical contributions to the mass \eqref{ms2}, compared to the negative bending contribution. It is assumed constant for simplicity, and lies in the range $-3 \leq \xi \leq 1$, so that $m_{s({\rm eff})}^2=(\xi+3)\etaperp^2 \geq 0$, ensuring a stable background. The phenomenology of interest in this paper is driven by the fact that a strong bending naturally generates a transient tachyonic instability of entropic and curvature perturbations.\footnote{This instability has been first noticed in \cite{Cremonini:2010ua}, and is similar in spirit to the amplification of gauge fields in axion-gauge field inflation \cite{Barnaby:2010vf}.}
 This is realized for $\xi<1$, but the range of scales that are amplified, as well as the magnitude of the enhancement, are strongly diminished as $\xi \to 1$. For $b \ll \etaperp^2$ as considered in Ref.~\cite{Fumagalli:2020adf}, this situation effectively corresponds to $\xi=0$, while $\xi=-3$ boils down to the setup of Ref.~\cite{Palma:2020ejf}.
The phenomenological model \eqref{eq:ms2-turn} thus encompasses and extends the two setups studied there.\\

\noindent {\bf Broad vs sharp turns.---} If the duration of the turn is sufficiently long that modes that are enhanced have already exited the Hubble radius at the end of the turn (or more accurately their effective sound horizon), the dynamics of the curvature perturbation can be described by the single-field effective theory with imaginary sound speed \cite{Garcia-Saenz:2018ifx,Garcia-Saenz:2018vqf,Fumagalli:2019noh}. The latter has a broad range of applicability, but in this context, it describes the transient tachyonic instability of fluctuations induced by the strongly non-geodesic motion and the negative entropic mass $m_s^2$, and its predictive power has been confirmed by explicit comparisons with first-principles numerical computations in the full multifield theory, for the power spectrum and the bispectrum \cite{Garcia-Saenz:2018ifx,Fumagalli:2019noh}, as well as with analytic solutions at the level of the two-field description \cite{Bjorkmo:2019qno}. In addition to the exponential amplification of the power spectrum compared to the baseline result $\Pzero$, the statistics of primordial fluctuations in these scenarios are characterized by a large enhancement of the bispectrum in flattened configurations \cite{Garcia-Saenz:2018vqf}, and an analogous  outcome holds for the trispectrum and all higher-order correlation functions \cite{Fumagalli:2019noh,Ferreira:2020qkf}. In this framework, the resulting peak of $\mathcal{P}_\zeta(k)$ is featureless, and interestingly, it can be narrower than what is allowed in single-field models \cite{Fumagalli:2020adf}\footnote{This can also happen for sharp turns \cite{Palma:2020ejf}.} (see for instance \cite{Byrnes:2018txb,Carrilho:2019oqg,Ozsoy:2019lyy} for studies of this single-field bound).\\

\begin{figure}[t]\begin{center}
	\includegraphics*[width=9cm]{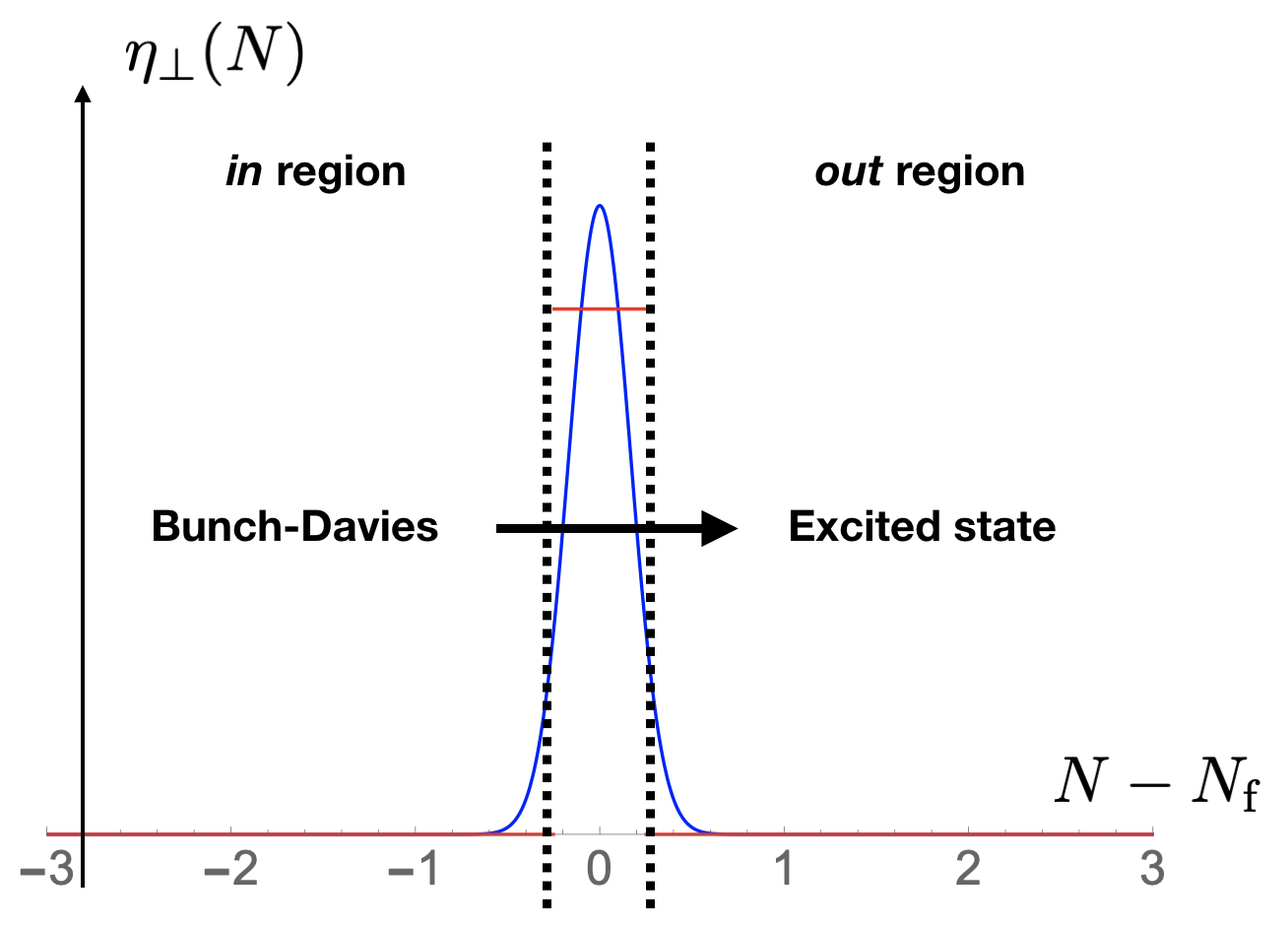}
	\caption{\textit{The strong and sharp turns which we concentrate on mostly affect scales that are sub-Hubble at the time of the turn: while in the Bunch-Davies vacuum in the \inr\,region before the turn, the latter effectively generates an excited state in the \outr\,region after the turn. This is responsible for the oscillations in the curvature power spectrum, and in the energy density power spectrum of the scalar-induced SGWB. The red and blue curves represent the time dependences of the bending parameter $\etaperp(N)$ for the top-hat and Gaussian profiles, Eqs.~\eqref{etaperp-top-hat} and \eqref{etaperp-Gaussian} respectively, with an arbitrary scale.
	}}
	\label{fig:setup}
	\end{center}
\end{figure}

In this paper, we will instead concentrate on sharp turns, with durations in \textit{e}-folds $\delta \ll 1$, and whose observational consequences are qualitatively different. In this case, the modes $k \sim \kstar$ of interest are still well inside the Hubble radius after the turn. The curvature perturbation is then decoupled from the entropic one, so that the dynamics of the former is single-field like and standard. However, the non-trivial evolution during the turn effectively generates an excited initial state for these modes (see fig.~\ref{fig:setup}), and as a consequence, in addition to an exponential enhancement, it results in the oscillations of the power spectrum ${\cal P}_\zeta(k)$ already noticed in fig.~\ref{fig:examples},  characteristic of sharp features, and which are here of order one.

\subsection{Enhancement of scalar fluctuations and a modulated amplitude from particle production}
\label{sec:enhancingP}

{\bf Sketch of the computation and result.---} Within the setup and assumptions described above, one can determine the power spectrum generated by sharp turns, generalising the computation of Ref.~\cite{Palma:2020ejf} (see also \cite{Bjorkmo:2019qno} for the case of broad turns, with the same parameterisation $m_s^2/H^2=(\xi-1)\etaperp^2$). 
Let us sketch here the computation. Essentially, it consists in matching the behaviours of the various operators describing the fluctuations, from the \inr\,region, where the Bunch-Davies vacuum is assumed, passing through the turn, and to the \outr\,region, resulting in an excited ``initial'' state, see fig.~\ref{fig:setup}. At the transitions between the regions, where $\etaperp$ is discontinuous, the continuity of the fields and of the momenta is used, in accordance with the equations of motion, and which ensures that the quantization conditions are duly preserved by the time evolution.\footnote{We thank Gonzalo Palma, Spyros Sypsas, Cristobal  Zenteno and Nicol\'as Parra for useful discussions on this point, as well as for sharing the details of their computation.} The behaviours in the \inr\,and \outr\,regions are standard, with each field dynamically decoupled from the other, while the turn in between is characterized by their exponential growth. Indeed, from the equations of motion deduced from the action \eqref{L2} (recall that one neglects the time-dependence of $H$ and $\epsilon$)

\begin{eqnarray}     \ddot{\zeta}_k+3H \dot{\zeta}_k+\frac{k^2}{a^2} \zeta_k&=&-\frac{2 \etaperp H}{\sqrt{2 \epsilon}}(\dot{Q}_{sk}+3H \dot{Q}_{sk})  \\
    \ddot{Q}_{sk}+3H \dot{Q}_{sk}+\left(\frac{k^2}{a^2}+m_s^2\right) Q_{sk} &=& 2 \etaperp H \sqrt{2 \epsilon} \dot{\zeta}_k
\end{eqnarray}
and neglecting the evolution of the scale factor during the short duration of the turn, one can look for solutions of the type $\zeta_k \propto e^{\pm i \omega_{\pm}\, t}$ and $Q_{s k} \propto e^{\pm i \omega_{\pm}\, t}$ (with each frequency $\omega_{\pm}$ appearing with both signs in the actual solution), finding \cite{Bjorkmo:2019qno}
\beq
\frac{\omega_{\pm}^2}{H^2}=\left(\frac{k}{\kf} \right)^2+\frac{3+\xi}{2}\etaperp^2 \pm \sqrt{4 \left(\frac{k}{\kf} \right)^2 \etaperp^2+\frac{(3+\xi)^2}{4} \etaperp^4 }\,.
\label{eq:frequencies}
\eeq
As anticipated, when the entropic field is tachyonic, corresponding to $\xi<1$, one finds that $\omega_-$ becomes imaginary for $k/\kf< \sqrt{1-\xi} \etaperp$, signaling the exponential growth of curvature and entropic fluctuations during the turn. After the turn (the \outr\,region in fig.~\ref{fig:setup}), and concentrating on the curvature perturbation $\zeta$ of observational interest, one can write on general grounds \begin{eqnarray} 
 \hat{\zeta}_k(\tau)&=& \bigg[ \alpha_k^\zeta\,\zeta^{\textrm{BD}}_k(\tau)  + \beta_k^\zeta \,\zeta^{*\textrm{BD}}_k(\tau) \bigg] \hat{a}_\zeta(\bk)+\textrm{h.c.}(-\bk) \nonumber \\
&+& \bigg[ \alpha_k^{\s}\,\zeta^{\textrm{BD}}_k(\tau)  + \beta_k^\s \,\zeta^{*\textrm{BD}}_k(\tau) \bigg] \hat{a}_\s(\bk)+\textrm{h.c.}(-\bk)\,,
\label{zeta-out-region}
\end{eqnarray}
where $\hat{a}_\zeta(\bk)$ and $\hat{a}_\s(\bk)$ are the operators annihilating the vacuum state of our multifield system $|0\rangle $, characterised by the absence of both ``$\zeta$ and $Q_s$ particles''. Here, 
\beq
\zeta^{\textrm{BD}}_k(\tau)= \left(\frac{k^3}{2 \pi^2}\right)^{-1/2} \Pzero^{1/2} e^{- i k \tau}(1+i k \tau)
\label{BD-mode-function}
\eeq
is the standard Bunch-Davies mode function expressed in conformal time $\tau$, and the quantisation condition 
imposes that the constant Bogolyubov coefficients verify 
\beq
\big(|\alpha_k^\zeta|^2-|\beta_k^\zeta|^2\big)+\big(|\alpha_k^\s|^2-|\beta_k^\s|^2\big)=1\,.
\label{eq:quantisation}
\eeq
The full expressions resulting from the computation, valid for all scales, can be found in appendix \ref{sec:analytics} but are not particularly illuminating. What is more interesting is to concentrate on scales $k/\kf < \sqrt{1-\xi} \etaperp$ that are exponentially amplified, and in particular scales of order $k \sim \kstar= \kf \etaperp$, for which 
the expressions simplify, up to an irrelevant $k$-dependent, but \textit{global} phase for all coefficients:
\begin{eqnarray} 
\label{alphas}
\alpha_k^\zeta&=&-\frac{e^{\deltaeta\, S}}{4 S \sqrt{1+\X}}\,, \quad  \alpha_k^\s=\frac{i e^{ \deltaeta\, S}}{4 S (1+\X+\sqrt{\X(1+\X)}} \, , \\
\beta_k^{\zeta,\s}&=& -\alpha_k^{\zeta,\s} \frac{S+i \x}{S-i \x}  e^{2 i e^{-\delta/2}  \x \etaperp} \, ,
\label{betas}
\end{eqnarray}
where 
\beq
\label{def-kappa-X}
\x = \frac{k}{\kstar} = \frac{k}{\kf \, \etaperp} \qquad \textrm{and} \qquad \X = \frac{{(3+\xi)}^2}{16 \x^2}
\eeq
and
\beq
\label{def-S}
S(k)=\sqrt{ \sqrt{4 \x^2 +\frac{(3+\xi)^2}{4}}-\left(\x^2 +\frac{(3+\xi)}{2} \right) 
}\,.
\eeq
We will discuss later in general terms the physical meaning of the simple expressions \eqref{alphas}-\eqref{betas}, but let us already quote the resulting expression for the power spectrum. Quite generally, from Eq.~\eqref{zeta-out-region}, it reads
\beq
\frac{{\cal P}_\zeta(k)}{\Pzero}= |\alpha_k^\zeta+\beta_k^\zeta|^2+|\alpha_k^\s+\beta_k^\s|^2\,,
\label{power-spectrum-formal}
\eeq
which gives here 
\beq
\label{eq:P-analytic-xi}
\frac{\mathcal{P}_\zeta(k)}{\Pzero} =   \, \frac{e^{2 \deltaeta\, S}}{2 S^2 \big(1 + \X + \sqrt{\X (1+\X)} \big)} \times \sin^2\left(e^{-\delta/2}\x \etaperp+\arctan(\x/S) \right)\,,
\eeq
again with domain of validity $\x \leq \sqrt{1-\xi}$. This is a simplified but very accurate expression that correctly reproduces the power spectrum for $k \sim \kstar$, i.e.~near the maximally enhanced scales. It is not meant to describe, neither smaller scales, nor larger ones. The figure~\ref{fig:examples} shows a comparison, for $\xi=-3$, between the numerically computed power spectrum, the analytical result based on the full coefficients \eqref{alpha-zeta}-\eqref{beta-s} (the two already successfully compared in Ref.~\cite{Palma:2020ejf} for $\xi=-3$), and the simple expression \eqref{eq:P-analytic-xi}, for different set of parameters. The agreement is  excellent irrespective of the value of $\xi$, as we will show later.\\

\noindent \textbf{Main features of the power spectrum.---} Given the good agreement between the numerical result for the scalar power spectrum and the analytical expression in \eqref{eq:P-analytic-xi}, we can focus on the latter to describe the main properties of $\mathcal{P}_\zeta(k)$. The analytical expression \eqref{eq:P-analytic-xi} shows that the power spectrum exhibits oscillations described by the $\sin^2$-term modulated by an envelope given by the function:
\begin{align}
\label{eq:P-envelope-xi} \frac{\mathcal{P}_{\textrm{env}}(k)}{\mathcal{P}_0} = \frac{e^{2 \deltaeta\, S}}{2 S^2 \big(1 + \X + \sqrt{\X (1+\X)} \big)} \, .
\end{align}
One important observation is that the envelope, as a function of $\x$, only depends on the model parameters $\delta$ and $\etaperp$ through the combination $\deltaeta$. The value of the envelope at its peak gives the maximal enhancement of the power spectrum compared to $\mathcal{P}_0$, For the expression \eqref{eq:P-envelope-xi} this will occur for a value $\x$ close to $\x_0=\tfrac{1}{4} \sqrt{7-6 \xi -\xi^2}$ where the function $S$ has its maximum.\footnote{It is not exactly at the maximum of $S$ due to the numerator in \eqref{eq:P-envelope-xi} which also makes it hard to extract the locus of the true maximum analytically.} There the scalar power spectrum is exponentially enhanced as $\mathcal{P}_\textrm{max} \sim \mathcal{P}_0 \exp \big(\tfrac{1}{2}(1-\xi) \deltaeta \big)$ where we have ignored non-exponential contributions. 

We now turn to the oscillations in $\mathcal{P}_\zeta(k)$. Firstly, by being proportional to $\sin^2$, the oscillations are $\mathcal{O}(1)$ in the sense that the power spectrum drops to zero between individual peaks.\footnote{This is a manifestation of the ``classical'' approximation here of considering $|\beta_k^{\zeta,s}|=|\alpha_k^{\zeta,s}|$. Rigorously, the power spectrum never drops to zero, but to values that are exponentially smaller than the one of the envelope.} Secondly, one can show that the peaks can be treated as periodic to a high degree of accuracy. To this end note that for $e^{-\delta / 2} \etaperp \gg \deltaeta$ the $\sin^2$-term changes much faster than the envelope (as a function of $\x$). This implies that the peaks of the power spectrum will coincide with the maxima of the $\sin^2$-term to a high degree of accuracy. In addition, for $e^{-\delta / 2} \etaperp \gg 1$ the term $e^{-\delta/2} \etaperp \x$ in the argument of $\sin^2$ changes much faster than the term $\arcsin(\x/S)$. Hence, as a first approximation the term $\arcsin(\x/S)$ can be treated as a constant phase compared to $e^{-\delta/2} \etaperp \x$. Combining the two observations it follows that the peaks are approximately periodic, with the periodic structure that of $\sin^2 \big(e^{-\delta/2} \x \etaperp \big)$. Thus the separation between peaks $\Delta \x$ can be approximated as 
\begin{align}
    \label{eq:xperiod}
    \Delta \x \approx \frac{\pi}{\etaperp} e^{\delta / 2} \, .
\end{align}
In $k$-space the period and the corresponding frequency can then be written as
\begin{align}
\label{eq:kperiod}
\Delta k \approx \pi e^{\delta / 2} \kf \, \quad \leftrightarrow \quad \omegalin=\frac{2\pi}{\Delta k}\approx\frac{2 e^{-\delta/2}}{\kf} \, ,
\end{align}
where have used $\x = k/ (\kf \, \etaperp)$ in \eqref{eq:xperiod}. The subscript on $\omegalin$ was included to indicate the linear spacing of peaks in $k$. Note that $\etaperp$ has dropped out from the expressions in \eqref{eq:kperiod} and it is $\kf$ that effectively sets the frequency of the oscillations in $k$-space.

We can also estimate the width of the individual peaks, which we define as the full width at half-maximum. As argued before, the oscillations in the power spectrum are well-approximated as $\sin^2$-peaks with argument $\pi \x /\Delta \x +\textrm{const}$. For $\sin^2$-peaks the full width at half maximum is just given by half the period. Thus, for the power spectrum \eqref{eq:P-analytic-xi} the width of the peaks in $\x$-space can be taken as $\Delta \x /2$.

The upshot is that the scalar power spectrum \eqref{eq:P-analytic-xi} can be approximately written in the schematic form
\begin{align}
\label{eq:P-sharp-template}
    \mathcal{P}_\zeta(k) \simeq \frac{1}{2} \mathcal{P}_{\textrm{env}}(k) \Big(1 + a \cos \big(\omegalin k + \varphi \big) \Big) \, , \quad \textrm{with} \quad a=1 \, ,
\end{align}
which can be identified as the template for a sharp feature with $\mathcal{O}(1)$ oscillations, see e.g.~\cite{Slosar:2019gvt}. \\

We can be more precise in the description of the scalar power spectrum for the special case $\xi=-3$, for which the expression \eqref{eq:P-analytic-xi} simplifies to
\begin{align}
\label{eq:P-analytic-xi-minus3} \frac{\mathcal{P}_\zeta(k)}{\mathcal{P}_0} \underset{\xi=-3}{=} e^{2 \sqrt{(2-\x)\x} \, \deltaeta} \, \frac{\Big( 1+ (\x-1) \cos \big(2 e^{-\delta / 2} \etaperp \x \big) + \sqrt{(2-\x)\x} \sin \big( 2 e^{-\delta / 2} \etaperp \x \big) \Big)}{4 (2-\x) \x} \, ,
\end{align}
with $\x \leq 2$. The envelope \eqref{eq:P-envelope-xi} becomes
\begin{align}
\label{eq:P-envelope-xi-minus3} \frac{\mathcal{P}_{\textrm{env}}(k)}{\mathcal{P}_0} \underset{\xi=-3}{=} \frac{e^{2 \sqrt{(2-\x)\x} \, \deltaeta}}{2 (2-\x) \x} \, . 
\end{align}
One upside compared to the case of general $\xi$ is that the maximum of the envelope \eqref{eq:P-envelope-xi-minus3} can be found exactly: The envelope peaks at $\x=1$, i.e.~$k=\kf \etaperp$, and the precise enhancement factor is given by $\mathcal{P}_{\textrm{max}} = \mathcal{P}_0 \exp(2 \deltaeta) /2$.
The periodicity of peaks can again be approximated by \eqref{eq:xperiod}. For the case $\xi=-3$ we can be more precise and compute corrections to the result in \eqref{eq:xperiod} for peaks near $\x=1$. This can be done systematically in an expansion in powers of $(\x-1)$ as is shown in appendix \ref{app:periodicity}. We find that at 1st order the effect is a constant shift in the period, while a drift in the period occurs only at 3rd order. A more accurate expression for the separation of peaks including the 1st order correction is then given in \eqref{eq:period-near-max-1st}.  

To summarise, the power spectrum as given in \eqref{eq:P-analytic-xi} exhibits an $\mathcal{O}(1)$ oscillation modulated by an exponentially enhanced envelope. The peaks can be treated as periodic in $k$-space to a high degree of accuracy with period \eqref{eq:kperiod}. In addition, the width of the peaks at half maximum is given by half the separation between neighbouring peaks.

\subsection{Boosted power spectrum from sharp features, and its inevitable order one rapid oscillations}
\label{sec:generic}

Using the example of the above computation, let us now interpret it and explain why its characteristic properties are generic to sharp features leading to a strong particle production and a boosted power spectrum, and hence why our results below concerning signatures in the stochastic background of gravitational waves are robust, and not tied to the specific scenario studied here. The genuine multifield aspects, i.e.~the fact that the curvature perturbation $\zeta$ in \eqref{zeta-out-region} is constructed out of different types of operators $\hat{a}_{\zeta,\s}(\bk)$, describing different excitations of the vacuum, is not important for this matter. Hence, to keep the discussion and notations simple, we will omit their indices $\zeta$ or $s$; what is more relevant indeed is that both types verify Eq.~\eqref{betas}, with clear physical origin and observational consequences. 

First, note that Eq.~\eqref{betas}, taken at face value, does not verify the quantisation condition \eqref{eq:quantisation}, as $|\beta|=|\alpha|$. This simply stems from the fact that the modes of interest are exponentially enhanced, and that we only kept (largely) dominant terms from the exact result \eqref{alpha-zeta}- \eqref{beta-s}, which itself does fulfill the quantisation condition. This is of course a manifestation that the quantum character is effectively lost because of a large particle production. Let us recall indeed that the Bunch-Davies mode function, with $\alpha=1$ and $\beta=0$, corresponds to the absence of particles, whereas an excited ``initial state'' of the type \eqref{zeta-out-region} corresponds to a non-zero occupation number $n_k=|\beta_k|^2$, which is exponentially greater than unity here. We can safely use the language of particles in this context, as the modes of interest are well inside the Hubble radius at the beginning of the \outr~region, corresponding to an unambiguous definition of the vacuum, and hence of particles.\\

The description with the excited state \eqref{zeta-out-region} in the \outr\, region is valid starting from the ``initial time'' $\Ni=\Nf+\frac{\delta}{2}$. The previous exponential growth of fluctuations implies that by the end of the turn, 
the mode function and its derivative are overwhelmly dominated by the real growing mode $\propto e^{-i \omega_{-}\,t}=e^{\etaperp S(k) N}$, and hence that both still share the same phase factor $e^{- i k \tau_{(\Nf-\delta/2)}}$ inherited from the Bunch-Davies behaviour in the \inr\,region. Hence, when doing the matching with the \outr\,region at time $\Ni$, 
and using $\zeta^{\textrm{BD}}_k \propto  k \tau e^{-i k \tau}$, one finds that $\alpha_k\, e^{-i k \tau_i}$ and $\beta_k\, e^{i k \tau_i}$ also approximately share the same phase, i.e.~that
\beq
\label{phase}
\textrm{ph} \left( \frac{\beta_k}{\alpha_k}\right) \sim e^{2 i k/\ki }\,,
\eeq
where $\ki$ is the scale crossing the Hubble radius at time $\Ni$, i.e.~with $-\ki \tau_i=1$, and where one can consider that $\alpha_k$ is real up to a global irrelevant phase factor. With $\ki=\kf\, e^{\delta/2}$, this correctly reproduces \eqref{betas}, up to a phase factor that varies comparatively much more slowly, and that this simple explanation cannot capture, but that, as we explained, plays basically no role for the maximally enhanced scales, and the overall observational implications. 
These explanations hopefully also justify why we kept the $e^{-\delta/2}$ term in \eqref{betas} and subsequent formulae: its appearance has a clear physical meaning, and is more robust than the precise assumptions $\delta \ll 1$ that led to the simple analytical results above. The exponential growth of the approximately real mode function is stopped at a particular time $\tau_i$, exciting the Bunch-Davies and non Bunch-Davies component at that time with the same phase, giving rise to oscillations in $k$-space with a frequency set by the scale $\ki$. In a related manner, we kept this factor in the benchmarks models that we used in the plots, with duration $\delta=\dea$ and $\delta=\deb$ that are not $\ll 1$.\\

What we have just described is not specific to our setup, but is characteristic of sharp features, that lead to a power spectrum (recall Eq.~\eqref{power-spectrum-formal})
\beq
\frac{{\cal P}(k)}{\Pzero} \sim |\alpha_k+\beta_k|^2 \sim |\alpha_k|^2+|\beta_k|^2+2 |\alpha_k| |\beta_k| \cos\left(\frac{2 k}{\kf}\right)\,,
\eeq
where again, we concentrate on the universal oscillations coming from \eqref{phase}, hence also with a scale dependence of $|\alpha_k|$ and $|\beta_k|$ much milder than the $\cos$ term, and we consider $\kf\simeq \ki$ to simplify the discussion here. Such oscillations, shared by higher-order correlation functions, have been extensively studied in different contexts and with various methods, see e.g. the reviews \cite{Chen:2010xka,Chluba:2015bqa,Slosar:2019gvt}.
However, most studies concentrate on CMB and LSS scales, hence they often have the observational and theoretical prejudice that the consequences of the feature should be mild enough to be in agreement with cosmological data indicating no evidence of a deviation from a smooth power-law spectrum (before or after Planck). Let us keep in mind that the quantisation condition imposes $|\alpha|^2-|\beta|^2=1$ (see Eq.~\eqref{eq:quantisation} in our setup), hence $|\beta|/|\alpha|$ can only range from $\ll 1$ to ${\cal O}(1)$. In all cases, the rough amplitude of the (envelope of the) power spectrum is $\mathcal{P}_{\textrm{env}} \sim \Pzero |\alpha|^2$, and hence 
\beq
{\cal P}(k) \sim \mathcal{P}_{\textrm{env}} \left( 1+\bigg|\frac{\beta}{\alpha}\bigg|^2+2 \bigg|\frac{\beta}{\alpha}\bigg| \cos\left(\frac{2 k}{\kf}\right)  \right)\,.
\eeq
This is compatible with a featureless power spectrum only for $|\beta| \ll |\alpha| \sim 1$, corresponding to a small amount of particle production. Of course, this type of small oscillations superimposed on an otherwise smooth power spectrum is extremely interesting, and may well exist and be detected in the future on CMB and LSS scales.

However, our emphasis in this paper is on smaller scales, for which observational constraints are much weaker, and perfectly allow even more dramatic consequences of features and particle production (the theoretical consistency of which is discussed in \eqref{sec:control}), in the range $|\beta| \sim |\alpha|$, leading to order one modulations of the power spectrum. Actually, with $\mathcal{P}_{\textrm{env}} \sim \Pzero |\alpha|^2$, all scenarios with a boosted power spectrum on small scales compared to the vanilla CMB amplitude $\Pzero \sim 10^{-9}$, necessarily have $|\alpha|^2 \gg 1$, and hence, with $|\alpha|^2-|\beta|^2=1$, they also have $|\beta|^2 \sim |\alpha|^2 \gg 1$, corresponding to a large particle production. Obviously, we are not saying that all models with a large bump of the power spectrum necessarily also have order one oscillations. However, the generic arguments above indicate that it is the case when the bump is created by a sharp feature. Moreover, as explained above, the particle production generated by sharp features concern modes that are well inside the Hubble radius at the time of the feature, i.e.~the relevant modes verify $k \gg \kf$. Hence, the order one modulation of the envelope, occurring with periodicity $\Delta k \sim \kf$, leads to many visible oscillations within the range of scales that are enhanced.\\

\begin{figure}[t]
\centering
\begin{subfigure}{.5\textwidth}
 \centering
   \begin{overpic}
[width=1.0\textwidth]{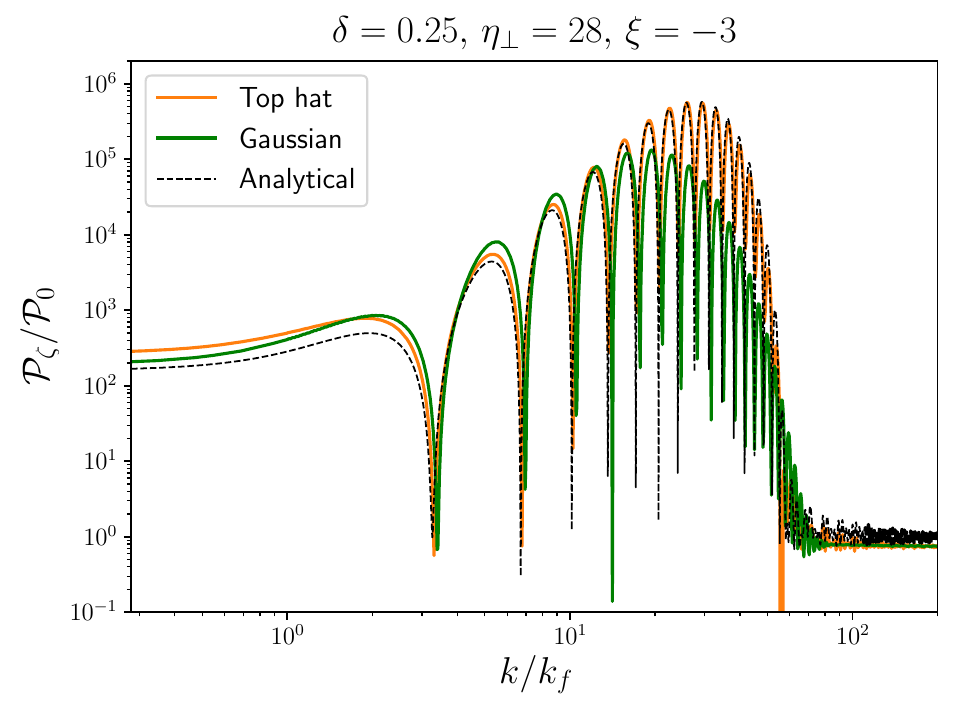}
\end{overpic}
\caption{\hphantom{A}}
\label{fig:figure2a}
\end{subfigure}%
\begin{subfigure}{.5\textwidth}
 \centering
   \begin{overpic}
[width=1.0\textwidth]{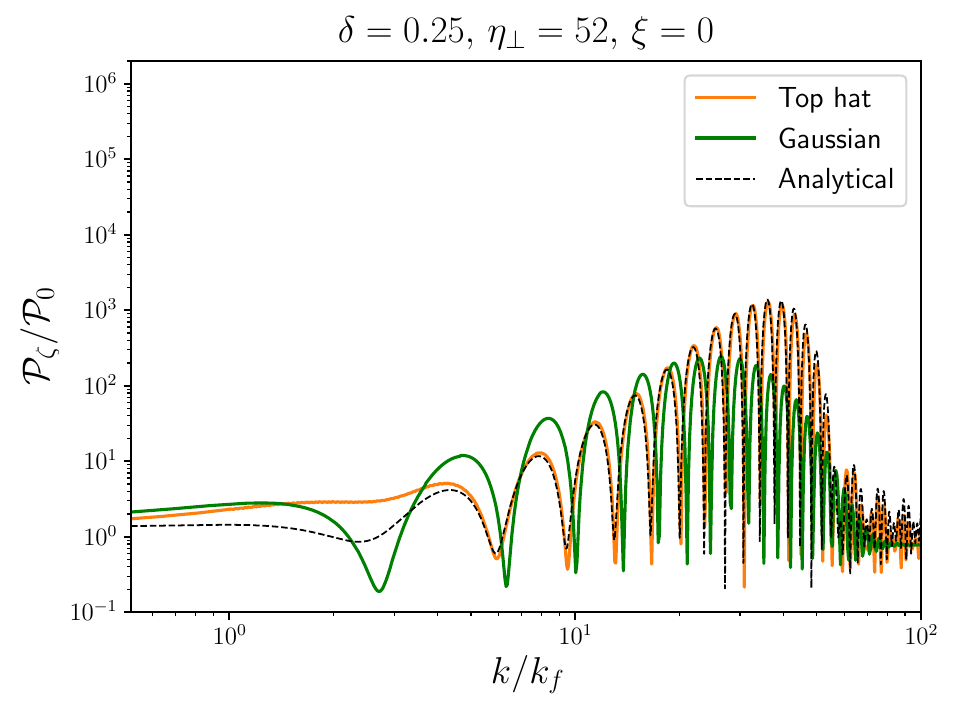}
\end{overpic}
\caption{\hphantom{A}}
\label{fig:figure2b}
\end{subfigure}%
\caption{\textit{The periodic modulations in $k$ of an exponentially enhanced envelope also holds for the Gaussian time-profile \eqref{etaperp-Gaussian}. This is shown here for $\xi=-3$ (left) and $\xi=0$ (right). In each case, we also plot the full numerical result for the top-hat profile, with the relationship \eqref{eq:top-hat-Gaussian-matching} between parameters for the two time-dependences, and the analytical result \eqref{eq:P-analytic-xi}. The excellent agreement numerical vs.~analytical already noticed for $\xi=-3$ holds for generic values of $\xi$, as shown by fig.(b) for $\xi=0$.}}
\label{fig:Gaussian-results}
\end{figure}

Given the genericity of the explanations that we just gave, it is no surprise that the power spectrum resulting from a sharp turn with a Gaussian time-dependence of $\etaperp(N)$ \eqref{etaperp-Gaussian} share the same characteristics as a sharp turn with a top-hat time-dependence: an exponentially enhanced envelope for scales $k \sim \kstar$, together with an order one rapid sinusoidal modulation of fixed frequency in $k$, see fig.~\eqref{fig:Gaussian-results} that includes a comparison with the top-hat profile, for $\xi=-3$ and $\xi=0$. Indeed, when the turn is sharp, it effectively selects a preferred time to stop the growth of the fluctuations, and hence it results in the $\cos(2k/\kf)$ modulations described above. Naturally, the precise value of $\kf$ may depend on the precise time-dependence of the bending parameter during the turn, but these are clearly secondary details, all the more as we have a complete freedom in choosing the time of the feature during inflation, at least at the level of this paper.

For completeness, we note that the frequencies of the oscillations for the two profiles \eqref{etaperp-Gaussian}-\eqref{etaperp-top-hat}, as well as the overall magnitude of the enhancement, are in good agreement if one chooses parameters such that
\beq
\sqrt{2 \pi} \frac{(\etamax)_{\textrm{Gaussian}}}{(\etamax)_{\textrm{top-hat}}} 	=\frac{\delta_{\textrm{top-hat}}}{\delta_{\textrm{Gaussian}}}=3\, ,
\label{eq:top-hat-Gaussian-matching}
\eeq
where the subscripts $\textrm{top-hat}$ and $\textrm{Gaussian}$ denote the values of the parameters assumed for each profile.
The parameters in fig.~\ref{fig:Gaussian-results} have been chosen to satisfy Eq.~\eqref{eq:top-hat-Gaussian-matching}.
The factor $3$ on the rhs is a numerical fudge factor that has been adjusted such that the frequencies agree. As for the first equality, it is simply such that the total ``angle'' swept by the turning trajectory $\Angle=\int \etaperp(N) d N$ is the same in both cases. This integrated bending is simply $\deltaeta$ in the top-hat model, and for the same reasons (almost by dimensionless analysis), it also controls the overall enhancement of the power spectrum for generic time-dependences (of course, given that the enhancement exponentially depends on $\Angle$, obtaining the same order of magnitude for the two time-dependences of $\etaperp(N)$ is already non-trivial).

\subsection{Backreaction and perturbativity}
\label{sec:control}

\textbf{Backreaction.---} Until now, we have assumed that the fluctuations evolve on a predetermined background. This is a legitimate assumption if the energy density $\rho $ of the particles created by the sharp feature is negligible compared to the background energy density  $\rhob=3 H^2 \Mp^2$, i.e.~if their backreaction is negligible. The energy density \beq
\rho=\frac{1}{a^4} \int \frac{\textrm{d}^3 \bk}{(2 \pi)^3} k\left( |\beta_k^\zeta|^2+|\beta_k^\s|^2 \right)
\eeq
of the relativistic particles that are created dilutes as $1/a^4$, hence the backreaction is most constraining at the beginning of the \outr\,region, and we evaluate it at that time in the following. Obviously, similar computations have already been performed in the general context of excited initial states in the literature. In particular, the generic estimates made in Ref.~\cite{Holman:2007na} also hold here, where, simply by dimensional analysis, the cutoff scale $M$ there is identified with $\Mn=H \etaperp$, and where the $|\beta|^2$'s for the relevant excited scales are of order the enhancement $\mathcal{P}_\textrm{max}/\Pzero$. This gives us the qualitative bound $\etaperp^4 \epsilon \mathcal{P}_\textrm{max} \lesssim 1$. This is confirmed by an explicit computation, which is not more difficult. With  $\mathcal{P}_{\textrm{env}}(k)/\Pzero=4 \left(|\beta_k^\zeta|^2+|\beta_k^\s|^2  \right) $, one obtains
\beq
\frac{\rho}{\rhob}=\frac{H^2}{24 \pi^2 \Mp^2} \etaperp^4 e^{-2 \delta} \int {\rm d} \x \x^3 \frac{\mathcal{P}_{\textrm{env}}(\x)}{\Pzero}\,.
\eeq
One can check that the dimensionless integral on the rhs is $\simeq 1/2 \,e^{\frac12 (1-\xi) \deltaeta}$ where the numerical factor $1/2$ is a proxi for a midly varying function of $\xi$ and $\deltaeta$, and where the exponential factor is simply $e^{2 \deltaeta\, S}$ at the maximum of $S$. This gives indeed the constraint
\beq
\frac{\rho}{\rhob}\simeq \frac{(1-\xi)}{12} \epsilon \etaperp^4 {\cal P}_{\textrm{max}} \lesssim 1\,.
\label{backreaction-result}
\eeq
An expected feature of this constraint, as well as other ones below, is that the large enhancement of the power spectrum $\mathcal{P}_\textrm{max}$, compared to conventional setups with ${\cal P} \sim 10^{-9}$, renders this bound comparatively more restrictive. For the orders of magnitudes $\etaperp \sim 10$ and $\epsilon \sim 10^{-2}$,
this gives the rough constraint ${\cal P}_{\textrm{max}} \lesssim 10^{-2}$.

In general, it is not enough to guarantee that the energy density of the produced particles is smaller than the background energy density. One must also ensure that this does not alter the almost de Sitter expansion, i.e.~that the contribution from the energy density and pressure of the particles to the slow-varying parameters $\epsilon$ and $\eta=\dot{\epsilon}/(H \epsilon)$ are negligible. Following the generic estimates of \cite{Holman:2007na}, this leads to the more stringent conditions $\etaperp^4 {\cal P}_\zeta \lesssim 1$ and $\etaperp^4 {\cal P}_\zeta \lesssim \eta$, respectively. In the following, we do not consider these constraints any further. If violated, one would have to take into account the temporary contribution from the particles created by the sharp feature, to $\epsilon(N)$ and its derivatives. However, by concentrating on the essential feature of a sharp strong turn, i.e.~a temporary boost of $\etaperp(N)$, while assuming a featureless expansion rate $H(N)$, we have since the start voluntarily kept a minimal number of ingredients, with the aim of highlighting in a simplified setup the observational consequences for SGWB of sharp features and copious particle production. Taking into account possible backreaction effects, after and during the turn, is very interesting, but is meaningful only if a more realistic description of the background and of the expansion rate is also envisaged in the first place, which we leave for future work.\\

\textbf{Perturbativity.---} On general physical grounds, it is well understood that one cannot consider arbitrary sharp features, since fluctuations would become strongly coupled and perturbative unitarity would be lost. In single-field setups with sharp features, perturbative unitarity bounds have been derived e.g.~in \cite{Bartolo:2013exa,Adshead:2014sga,Cannone:2014qna}. There, it has been found that even a modest, but sharp feature in the expansion rate, of duration $\delta$ \textit{e}-folds, boosts interactions at any order in fluctuations, rendering the bounds ${\cal L}_n/{\cal L}_2 \lesssim 1$ at large $n$ more stringent than simply ensuring it for $n=3$, imposing the constraint ${\cal P}_\zeta/\delta^4 \lesssim 1$ (for a sound speed $c_s=1$). Our setup is different, but we will see that preliminary estimates give similar results. Two additional (and related) sources of complexity are present here: the transient tachyonic instability makes the use of standard perturbative unitarity bounds questionable, and the resulting large amplitude of the curvature fluctuation is a threat by itself to a perturbative expansion.

The latter point was addressed in Ref.~\cite{Palma:2020ejf}, where it was found that ${\cal P}_\zeta \ll 1$ was enough to warrant perturbative control, based on the criterion ${\cal L}_3/{\cal L}_2 \ll 1$. More refined estimates indicate that the constraint from perturbative control is more severe. The fact that the large parameter $\etaperp$ comes exclusively from the kinetic term in the action results in a particular structure of the schematic form ${\cal L}_n/{\cal L}_2 \sim (\dot{\zeta}/H)^{n-2}$, as used in \cite{Palma:2020ejf} for $n=3$.
However, in such models with sharp features, the dominant interactions arise not at Hubble crossing for the relevant modes, corresponding to energies of order $H$, but around the time of the feature, inside the Hubble radius, and corresponding to energies of order $\Mn=H \etaperp$. Hence, in order to ensure perturbative control, the ratios ${\cal L}_n/{\cal L}_2$ should be evaluated at that scale (see e.g.~\cite{Behbahani:2011it,Bartolo:2013exa}). With $\dot{\zeta}/H_{|\Mn} \sim \etaperp \zeta_{\Mn}$, corresponding to the growth of fluctuations $\propto e^{|\omega_{-}|t}$, and with $\zeta_{\Mn} \sim \frac{\Mn}{H} \zeta_H$, corresponding to the sub-Hubble behaviour in the \outr\,region, where $\zeta_H$ is the final observable value of the curvature perturbation, one gets the constraint
\beq
\frac{{\cal L}_n}{{\cal L}_2} \bigg|_{\Mn} \sim \left(\etaperp^2 \sqrt{{\cal P}_\zeta}\right)^{n-2} \lesssim 1, 
\label{Ln}
\eeq
i.e.~$\etaperp^4 {\cal P}_{\textrm{max}} \lesssim 1$. This is more stringent than the backreaction bound \eqref{backreaction-result}, and more stringent than just ${\cal P}_{\textrm{max}} \ll 1$, coming from the fact that the main interactions arise on sub-Hubble scales. Interestingly, as the amplitude of the power spectrum is determined by the combination $\deltaeta$, for fixed ${\cal P}_{\textrm{max}}$, the above constraint is similar to the one mentioned above. That is, how strong the turn is (and hence how deep inside the Hubble radius are the scales affected by the feature), and how sharp the turn is, quantified respectively by $\etaperp$ and $1/\delta$, affect the bounds ensuring theoretical controls in a qualitatively similar way. 
When perturbativity is guaranteed, the estimate \eqref{Ln} indicates a non-Gaussianity parameter $f_{\textrm{NL}} \sim \etaperp^2$ for generic triangular configurations, although as usual with excited initial states, one also expects an additional enhancement in flattened configurations. Concerning perturbative control, these back-of-the-envelope estimates are a proxy for a proper computation of the one-loop contribution to the power spectrum, which would provide a quantitative criterion, but which lies outside the scope of this work.\\

For the rest of the paper, considering a fiducial value $\epsilon \sim 10^{-2}$ (naturally, the bound from backreaction becomes less stringent if $\epsilon$ assumes smaller values), we summarise the tentative theoretical constraints discussed above as:
\begin{eqnarray} 
\label{backreaction}
\textrm{No backreaction:} \qquad &&\etaperp^4 e^{\frac12(1-\xi) \deltaeta} \lesssim 10^{11} \left(\frac{10^{-9}}{\Pzero}\right) \, ,
\\
\textrm{Perturbative control:} \qquad &&\etaperp^4 e^{\frac12(1-\xi) \deltaeta} \lesssim 10^{9} \left(\frac{10^{-9}}{\Pzero}\right)\,.
\label{perturbativity}
\end{eqnarray}
Contrary to what these equations may suggest, considering values of $\Pzero >10^{-9}$ (moderately) alleviates these bounds: with a greater baseline power spectrum $\Pzero$, generating a given amplitude of the curvature power spectrum necessitates a comparatively smaller values of the bending parameter $\etaperp$, hence it is easier to satisfy the bound \eqref{Ln}. It is likely that further studies dedicated to rigorously ensuring perturbative control, taking into account the backreaction of the particles produced during the turn, as well as more realistic description of the turn, may reduce the allowed parameter space, as well as give interesting twists to the phenomenology described in this paper. However, we expect our main results to be robust, and we think that these questions open several interesting avenues of research. Keeping this in mind, in the following, we study the consequences for the SGWB of an enhanced and oscillatory curvature power spectrum of the type described in section \ref{sec:enhancingP}.

\section{Scalar-induced stochastic background of gravitational waves}
\label{sec:stochGWs}
In this section we will compute the contribution to the SGWB induced by the scalar fluctuations sourced during inflation. The contribution studied here is produced as the scalar fluctuations re-enter the horizon after the end of inflation, sourcing GWs at second order in perturbations. In this case the fraction of energy density in gravitational waves can be computed in an essentially model-independent fashion, depending on the power spectrum of scalar fluctuations and the equation of state of the universe at the time of re-entry \cite{Baumann:2007zm}.\footnote{The connected four-point function of the curvature perturbation, i.e.~the primordial trispectrum, can also in general impact the resulting GW spectrum \cite{Unal:2018yaa,Atal:2021jyo}. We do not discuss here this model-dependent contribution.} Here, to be specific, we assume radiation-domination (RD) during the phase of re-entry of the relevant (enhanced) scalar fluctuations. The fraction of energy density in gravitational waves today is then given by \cite{Baumann:2007zm,Espinosa:2018eve}:
\begin{align}
\label{eq:OmegaGW-i}
    \Omega_{\textrm{GW}}(k) = c_g \Omega_{\textrm{r},0} \int_0^{\frac{1}{\sqrt{3}}} \textrm{d} d \int_{\frac{1}{\sqrt{3}}}^\infty \textrm{d} s \, \mathcal{T}_{\textrm{RD}}(d,s) \, \mathcal{P}_\zeta \bigg(\frac{\sqrt{3}k}{2}(s+d)\bigg) \mathcal{P}_\zeta \bigg(\frac{\sqrt{3}k}{2}(s-d)\bigg) \, ,
\end{align}
where $\Omega_{\textrm{r},0}$ is the fraction of energy density in radiation today and $\mathcal{T}_{\textrm{RD}}(d,s)$ is the integration kernel appropriate for GW production during RD\footnote{For GWs induced by scalar fluctuations during a period of matter-domination the corresponding integration kernel $\mathcal{T}_{\textrm{MD}}(d,s)$ is derived in \cite{Kohri:2018awv}.}:
\begin{align}
\label{eq:TRDds}
\mathcal{T}_{\textrm{RD}}(d,s) = 36 \frac{{\big(d^2-\tfrac{1}{3}\big)}^2{\big(s^2-\tfrac{1}{3}\big)}^2{\big(d^2+s^2-2\big)}^4}{{\big(s^2-d^2\big)}^8} \bigg[ \bigg(\ln \frac{1-d^2}{|s^2-1|} + \frac{2 \big(s^2-d^2\big)}{d^2+s^2-2}\bigg)^2 + \pi^2 \Theta(s-1) \bigg] .
\end{align}
The factor $c_g$ relates the energy density fraction in GW during RD to that today:
\begin{align}
    c_g \equiv \frac{a_{\textsc{rd}}^4 \, \rho_{\textrm{r},\textsc{rd}}}{a_{0}^4 \, \rho_{\textrm{r},0}} = \frac{g_{*,\textsc{rd}}}{g_{*,0}} {\bigg( \frac{g_{*S,0}}{g_{*S,\textsc{rd}}} \bigg)}^{4/3} \, ,
\end{align}
with $\rho_{\textrm{r}}$ the energy density in radiation. Assuming that the matter content of the universe is given by that of the Standard Model and that all degrees of freedom are relativistic when the enhanced scalar fluctuations re-enter, one finds $c_g \approx 0.4$. Then, the combination $c_g \Omega_{\textrm{r},0} h^2$ that will appear in the subsequent quantitative results is given by $c_g \Omega_{\textrm{r},0} h^2 = 1.6 \cdot 10^{-5}$.

\subsection{Characteristics of the scalar-induced gravitational wave signal}
\label{sec:GWproperties}
As reviewed in sec.~\ref{sec:generic}, particle production resulting from a sharp feature generically produces a scalar power spectrum with sinusoidal oscillations in $k$, modulated by a comparatively slowly varying envelope, whose form is model-dependent. This implies the following properties of the scalar power spectrum, which will be important for the subsequent analysis, and which can be also obtained in more general frameworks:
\begin{enumerate}
    \item \emph{The scalar power spectrum consists of a series of individual peaks that are periodic in $k$.}
\end{enumerate}
As soon as particle production is efficient, with occupation numbers $n_k \sim 1$ or larger, we have that:
\begin{enumerate}
\setcounter{enumi}{1}
\item \emph{The amplitude of modulations of the scalar power spectrum is order one.}\footnote{The subsequent analysis presented in this section will assume $\mathcal{O}(1)$ modulations of $\mathcal{P}_\zeta(k)$. At the end of section \ref{sec:Obs3} we comment on how our findings are changed if the amplitude of modulations is modified.}
\end{enumerate}
Moreover, when particle production is copious, with $n_k \gg 1$, an additional property is verified:
\begin{enumerate}
\setcounter{enumi}{2}
    \item \emph{The amplitude of the peaks is modulated by an envelope with an exponentially enhanced peak.}
\end{enumerate}
The converse is also true, i.e.~an exponential enhanced peak driven by a sharp feature comes inevitably with copious particle production and thus with $\mathcal{O}(1)$ oscillations in the power spectrum.

In the following, all three properties will play an important role in our analysis. For definiteness, we formulate our results with the specific example of the strong sharp turn presented in section \ref{sec:model}. However, the novel and most interesting features of the SGWB that we will discuss derive from property 1. Hence our results have a broad range of applicability to sharp features in general. Point 2 is important for maximising the detectability of the novel signal. The properties of the envelope play an interesting, but secondary role. 

\begin{figure}[t]
\centering
\begin{overpic}[width=1.0\textwidth]{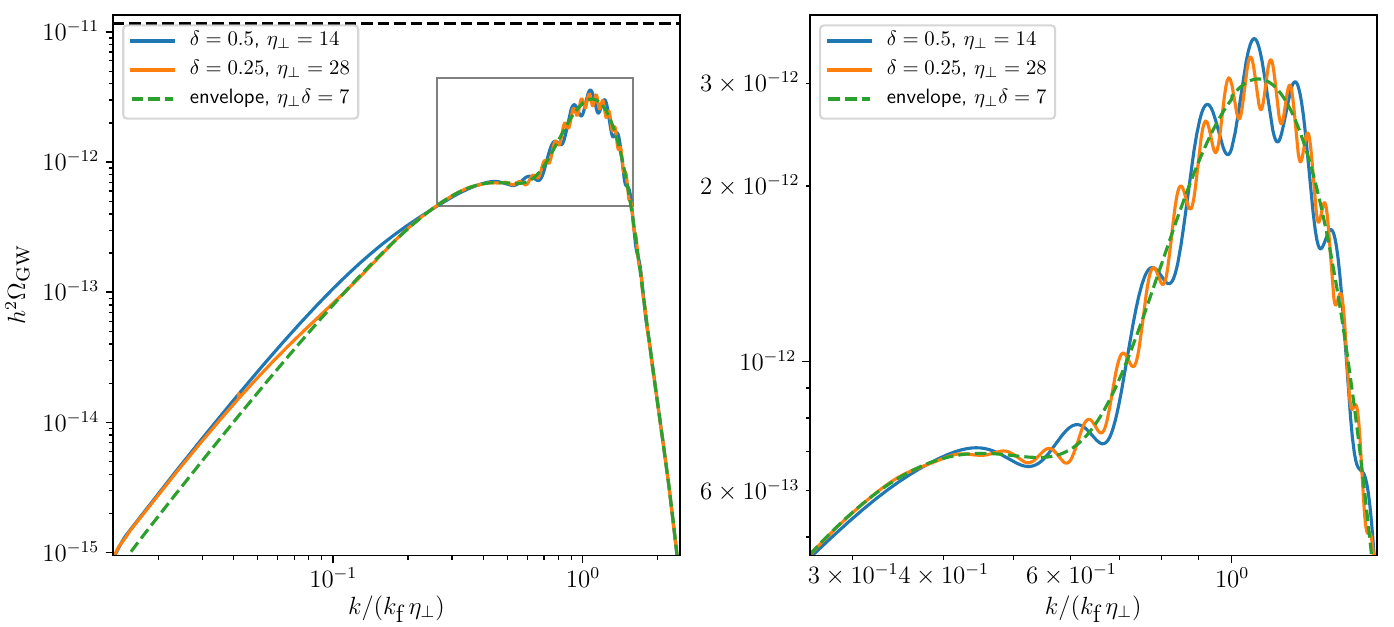}
\end{overpic}
\caption{\textit{$h^2 \Omega_{\textrm{GW}}(k)$ computed from \eqref{eq:OmegaGW-i} for the scalar power spectrum \eqref{eq:P-analytic-xi-minus3}. The blue plot is for the parameter choices $(\delta, \etaperp)=(\deA,\etaA)$ and the orange plot for $(\deB,\etaB)$. The green dashed line depicts $h^2\Omega_{\textrm{GW}}(k)$ computed for the envelope of the power spectrum given in \eqref{eq:P-envelope-xi-minus3}. The plot on the RHS depicts the region the gray box in the LHS plot in greater detail. The black dashed line on the LHS is the estimate \eqref{eq:Omega-max-result} for the max.~amplitude of $h^2 \Omega_{\textrm{GW}}$.}}
\label{fig:OmegaGWofx_deltaetaperp}
\end{figure}

The scalar power spectra considered here consist of a series of individual peaks. GW signals induced by scalar perturbations with such a multi-peaked structure have been studied in \cite{Cai:2019amo} with a focus on series of $\delta$-function and Gaussian peaks. While the findings in \cite{Cai:2019amo} will be important for our work, we will find that the periodic structure of peaks and the existence of an exponentially enhanced envelope give rise to new characteristic features in the SGWB that have not been discussed before.

Our findings are best illustrated by considering explicit examples. Thus, in the following we will focus on the SGWB signal as produced by scalar fluctuations with a power spectrum as given in \eqref{eq:P-analytic-xi-minus3}, i.e.~for a sharp turn during inflation with a top-hat profile in time with $\xi=-3$. This scalar power spectrum exhibits all the properties listed above and is hence ideal for illustration.  To show quantitative results in the plots, the normalisation $\mathcal{P}_0=\PCMB$ is assumed.

We will use as benchmark models the ones whose scalar power spectra have already been show in fig.~\ref{fig:examples}, with model parameters $(\delta, \etaperp)=(\deA,\etaA)$ and $(\deB,\etaB)$. The two example models share the same value of the product $\deltaeta$ and as a result their power spectra have a common envelope \eqref{eq:P-envelope-xi-minus3} when viewed as a function of $\x= k / (\kf \etaperp)$.
The corresponding GW energy density power spectra $h^2 \Omega_{\textrm{GW}}(k)$, computed via \eqref{eq:OmegaGW-i}, are shown in fig.~\ref{fig:OmegaGWofx_deltaetaperp}.
The plots of $h^2 \Omega_{\textrm{GW}}(k)$ for the two example models exhibit a broad lower peak for lower values of $k$ and a narrower higher principal peak for larger values of $k$ separated by a shallow dip. The GW signal then drops off steeply for $k / (\kf \etaperp) \gtrsim 2$. The higher peak exhibits modulations which differ between the models with different values of $(\delta, \etaperp)$. In the following we will describe the main features of this GW signal in more detail and relate it to the properties of $\mathcal{P}_\zeta(k)$.

\subsubsection{The overall shape of $\Omega_{\textrm{GW}}(k)$ is determined by the envelope of the scalar power spectrum}
\label{sec:Obs1}
This is supported by the observation that, up to the modulations on the main peak which will be discussed in detail later, the two examples give rise to GW spectra that closely trace one another (as functions of $\x= k / (\kf \etaperp)$). Further evidence is given by the GW signal (see the green dashed line) computed for the envelope \eqref{eq:P-envelope-xi-minus3} of the two power spectra: Over the broad left peak this is near-indistinguishable from the GW signal of the two example models and behaves like a smoothed version along the higher narrow peak. 

To be precise, what is plotted as the green dashed curve in fig.~\ref{fig:OmegaGWofx_deltaetaperp} is $h^2 \Omega_{\textrm{GW}}(k)$ as computed via \eqref{eq:OmegaGW-i} for the envelope of the power spectrum \eqref{eq:P-envelope-xi-minus3} multiplied by an additional normalisation factor $1/ \mathcal{N}$. This factor accounts for the fact that the envelope of the power spectrum in \eqref{eq:P-envelope-xi-minus3}, lacking the oscillations of the full expression \eqref{eq:P-analytic-xi-minus3}, possesses a larger total power (i.e.~integral over $k$) than the actual power spectrum. The GW energy density power spectrum $\Omega_{\textrm{GW}}$, as given in \eqref{eq:OmegaGW-i}, contains two integrations over two factors of the scalar power spectrum, which suggests that $\Omega_{\textrm{GW}}$ is sensitive to the square of the total power in $\mathcal{P}_\zeta(k)$. This is what we assume in the following. In this case, for \eqref{eq:OmegaGW-i} computed for the envelope \eqref{eq:P-envelope-xi-minus3} to not only reproduce the overall shape but also the correct amplitude of the GW signal of the full power spectrum \eqref{eq:P-analytic-xi-minus3}, the expected normalisation factor is:
\begin{align}
\label{eq:GW-envelope-normalisation}
\frac{1}{\mathcal{N}} = {\left( \frac{\int_0^\infty \textrm{d} k \, \mathcal{P}_\zeta(k)}{\int_0^\infty \textrm{d} k \, \mathcal{P}_{\textrm{env}}(k)} \right)}^2 \, . 
\end{align}
For the power spectrum \eqref{eq:P-analytic-xi-minus3} with envelope \eqref{eq:P-envelope-xi-minus3} one finds that to a very good approximation $\int_0^\infty d k \, \mathcal{P}_{\textrm{env}}(k) \simeq 2 \int_0^\infty d k \, \mathcal{P}_\zeta(k)$ and hence $1/ \mathcal{N} \simeq 1/4$. This is what has been used in fig.~\ref{fig:OmegaGWofx_deltaetaperp} and we take the close match in amplitude between the various results for $\Omega_{\textrm{GW}}$ as evidence for our assumption that $\Omega_{\textrm{GW}}$ is sensitive to the square of the total power in $\mathcal{P}_\zeta(k)$.\footnote{An alternative definition of the normalisation factor $1/\mathcal{N}$ is in terms of the ratio of variances, $1/\mathcal{N} = \big(\langle \zeta^2 \rangle / \langle \zeta^2 \rangle_{\textrm{env}} \big)^2$, with $\langle \zeta^2 \rangle= \int_{-\infty}^{\infty} \textrm{d} \ln k \, \mathcal{P}_\zeta(k)$ and $\langle \zeta^2 \rangle_{\textrm{env}}= \int_{-\infty}^{\infty} \textrm{d} \ln k \, \mathcal{P}_\textrm{env}(k)$. The upside of this definition is that this gives $1/\mathcal{N}$ in terms of quantities with clear physical meaning. As the power spectrum is highly peaked over a small range of $k$-values the normalisation thus defined will not differ much numerically from that given in \eqref{eq:GW-envelope-normalisation}.} 

\begin{figure}[t]
\centering
\begin{subfigure}{.5\textwidth}
 \centering
   \begin{overpic}
[width=1.0\textwidth]{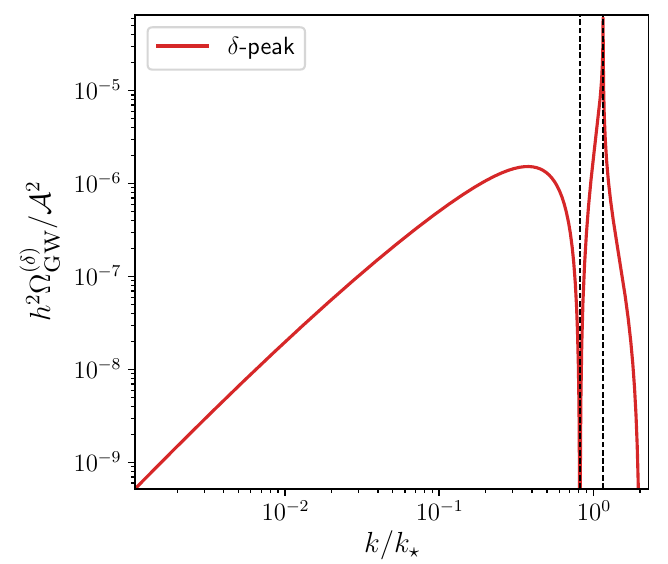}
\end{overpic}
\caption{\hphantom{A}}
\label{fig:Omega-delta-plot}
\end{subfigure}%
\begin{subfigure}{.5\textwidth}
 \centering
   \begin{overpic}
[width=1.0\textwidth]{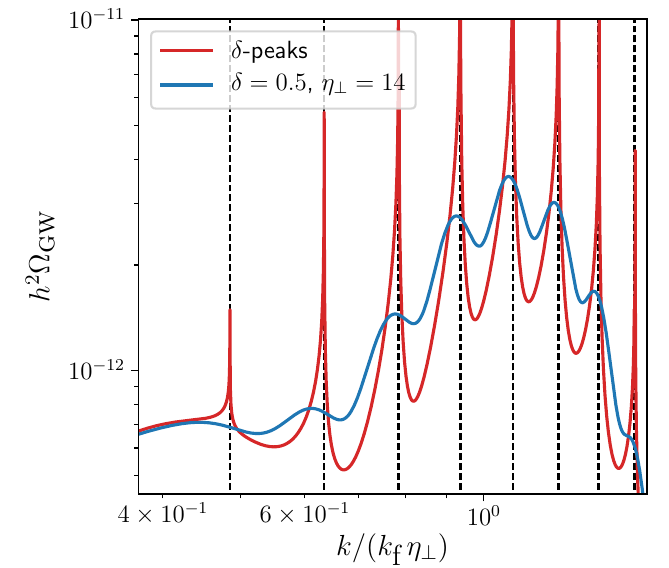}
\end{overpic}
\caption{\hphantom{A}}
\label{fig:Omega-delta-etap-vlines}
\end{subfigure}%
\caption{\textbf{(a):} \textit{$h^2 \Omega_{GW}^{(\delta)}(k)$ in \eqref{eq:OmegaGW-i-delta} induced by a $\delta$-peak scalar power spectrum \eqref{eq:P-delta}. The dashed vertical lines denote the loci $k = \sqrt{2/3} k_\star$ where $\Omega_{GW}^{(\delta)}$ vanishes and $k = 2/\sqrt{3} k_\star$ where $\Omega_{GW}^{(\delta)}$ diverges.} \textbf{(b):} \textit{$h^2 \Omega_{GW}(k)$ (blue curve) for the model with $(\delta, \etaperp)=(\deA,\etaA)$. The red curve shows $h^2 \Omega_{\textrm{GW}}^{(\delta, n_\textrm{p})}(k)$, i.e.~the GW spectrum
in \eqref{eq:Omega-from-delta-comb} for the $\delta$-peak model of the power spectrum given in \eqref{eq:P-as-delta-comb}. The dashed vertical lines denote the expected locations of peaks as predicted by the $\delta$-peak model.}}
\label{fig:deltapeaks}
\end{figure}

To discuss the overall spectral shape of the SGWB, we can hence ignore the individual peaks in the scalar power spectrum and just focus on the envelope. The shape of $\Omega_{\textrm{GW}}(k)$ observed here, i.e.~the appearance of a broad lower peak and a narrower higher peak separated by a dip, can be identified as characteristic for a scalar power spectrum that exhibits a single sufficiently narrow peak, as has been discussed for Gaussian-peaked scalar power spectra in \cite{Cai:2019amo}. We can make contact with this analysis by observing that the envelope of the power spectrum $\mathcal{P}_\textrm{env}(k)$ as given in \eqref{eq:P-envelope-xi-minus3} is well-approximated by a Gaussian peak as
\begin{align}
    \label{eq:env-Gauss-match}
    \mathcal{P}_{\textrm{env}} (k) \approx \mathcal{P}_{\textrm{env}}(\kstar) \, \exp \bigg(- \frac{(k-\kstar)^2}{2 \sigma_{\textrm{env}}^2} \bigg) \, , \quad \textrm{with} \quad \sigma_{\textrm{env}} = \frac{\kstar}{\sqrt{2(\deltaeta-1)}} \, ,
\end{align}
and $\kstar = \kf \, \etaperp$. For strong sharp turns we have $\deltaeta \gg 1$ and hence $\kstar \gg \sigma_{\textrm{env}}$, i.e.~the Gaussian peak is narrow. For a narrow-peaked scalar power spectrum the corresponding GW fraction shares many properties with that obtained for a scalar power spectrum given by a single $\delta$-function peak, which can be interpreted as the extreme limit of a narrow-peaked scalar power spectrum. To be specific, consider a power spectrum exhibiting a single $\delta$-peak at $k=\kstar$:
\begin{align}
\label{eq:P-delta}
    \mathcal{P}_\zeta^{(\delta)}(k) = \mathcal{A} \, \delta (k-\kstar ) \, .
\end{align}
Inserting this into \eqref{eq:OmegaGW-i} the corresponding expression for $\Omega_{\textrm{GW}}$ can be computed analytically \cite{Kohri:2018awv}, finding:
\begin{align}
\nonumber \Omega_{\textrm{GW}}^{(\delta)}(k) &= c_g \Omega_{\textrm{r},0} \, \frac{\mathcal{A}^2}{3k^2} \mathcal{T}_{\textrm{RD}} \Big(0, \tfrac{2\kstar}{\sqrt{3}k} \Big) \, \Theta\Big(2 -\tfrac{k}{\kstar} \Big) \\ 
\nonumber &= c_g \Omega_{\textrm{r},0} \frac{3 \mathcal{A}^2 k^2}{4 \kstar^4} {\Big(1 - \tfrac{k^2}{4 \kstar^2} \Big)}^2 {\Big(1 - \tfrac{3 k^2}{2\kstar^2} \Big)}^4 \times \\
\label{eq:OmegaGW-i-delta} & \hphantom{= c_g \Omega_{\textrm{r},0}} \times \bigg[{\bigg( \ln \Big| 1 - \tfrac{4\kstar^2}{3k^2} \Big| - \frac{2}{1 - \tfrac{3 k^2}{2\kstar^2}} \bigg)}^2 + \pi^2 \Theta\Big(2 - \sqrt{3}\tfrac{k}{\kstar} \Big) \bigg] \Theta\Big(2 -\tfrac{k}{\kstar} \Big) \, .
\end{align}
A plot of $\Omega_{\textrm{GW}}^{(\delta)}(k)$ is shown in fig.~\ref{fig:Omega-delta-plot}.
This exhibits a broad peak at $k \sim \kstar/ e$, a dip at $k \sim \sqrt{2/3} \kstar$ where $\Omega_{\textrm{GW}}^{(\delta)}(k)$ vanishes and a principal peak that diverges at $k=2/\sqrt{3} \kstar$ due to resonant amplification. A scalar power spectrum with a finite-width peak that is sufficiently narrow will inherit this structure with the exception that at the dip the GW fraction does not drop all the way to zero and the amplitude of the principle peak is finite \cite{Pi:2020otn}.\footnote{The broader the peak in $\mathcal{P}_\zeta(k)$ the more the two peaks in $\Omega_{\textrm{GW}}$ merge into one, eliminating the intervening dip. See e.g~\cite{Pi:2020otn} where this behaviour is analysed for a log-normal peak of varying width in $\mathcal{P}_\zeta(k)$.} This is indeed what we observe here for the overall shape of $\Omega_{\textrm{GW}}(k)$ which can thus be understood as a direct consequence of the fact that $\mathcal{P}_\textrm{env}(k)$ is narrow-peaked.

Next we will provide an estimate for the amplitude of the principle peak of $\Omega_{\textrm{GW}}$. Following \cite{Pi:2020otn}, for a narrow-peaked scalar power spectrum the amplitude of $\Omega_{\textrm{GW}}$ at the principal peak can be estimated by taking the result for a $\delta$-peak and smoothing $\Omega_{\textrm{GW}}^{(\delta)}$ over an appropriate scale. A natural smoothing scale is given by the width of the peak in the scalar power spectrum \cite{Pi:2020otn}, which in our case will be taken as $2 \sigma_\textrm{env}$. Smoothing is then done by averaging over an interval $2 \sigma_\textrm{env}$ centred at the value of the resonant peak at $k = 2 / \sqrt{3} \kstar$:
\begin{align}
\label{eq:Omega-max-smoothing}
    \Omega_{\textrm{GW}}^{\textrm{max}} = \frac{1}{2 \sigma_{\textrm{env}}} \int_{\frac{2}{\sqrt{3}}\kstar-\sigma_{\textrm{env}}}^{\frac{2}{\sqrt{3}}\kstar+\sigma_{\textrm{env}}} \textrm{d}k \, \Omega_\textrm{GW}^{(\delta)} \, ,
\end{align}
To perform this calculation we need to specify a value for $\mathcal{A}$ in \eqref{eq:OmegaGW-i-delta}. From \eqref{eq:P-delta} it follows that $\mathcal{A}$ corresponds to the total power in the scalar power spectrum, which for our Gaussian fit \eqref{eq:env-Gauss-match} is given by
\begin{align}
    \mathcal{A} = \frac{\sqrt{2 \pi} \sigma_{\textrm{env}} \mathcal{P}_{\textrm{env}}(\kstar)}{\sqrt{\mathcal{N}}} = \frac{\sqrt{ \pi} \, \kstar \mathcal{P}_0 \, e^{2 \deltaeta}}{\sqrt{\mathcal{N}} \, \sqrt{\deltaeta -1}} \, ,
\end{align}
where we also included the required normalisation factor $1/ \sqrt{\mathcal{N}}$. The integration in \eqref{eq:Omega-max-smoothing} can then be done explicitly to find:
\begin{align}
\label{eq:Omega-max-result}
\Omega_{\textrm{GW}}^{\textrm{max}} \approx \frac{\pi}{9 \mathcal{N}} c_g \Omega_{r,0} \, \frac{\mathcal{P}_0^2 \, e^{4 \deltaeta}}{(\deltaeta -1)} \bigg[{\bigg( \frac{1}{2} \log \bigg(\frac{3}{2 (\deltaeta -1)} \bigg) +1 \bigg)}^2 + 1+ \frac{\pi^2}{2} \bigg] \, .
\end{align}
For the examples shown in fig.~\ref{fig:OmegaGWofx_deltaetaperp} we plotted $\Omega_{\textrm{GW}}^{\textrm{max}}$ as the dashed horizontal line, finding that this reproduces the maximal amplitude of the principal peak up to an $\mathcal{O}(1)$ factor, overestimating the amplitude by a factor of $\sim 4$. 

To summarise, for a scalar power spectrum consisting of individual peaks with a periodic structure, the overall shape of the resulting scalar-induced GW signal is determined by the corresponding GW signal for the envelope of the scalar power spectrum (with appropriate normalisation). If this envelope is sufficiently narrowly peaked about its maximum at $k=\kstar$ the overall shape of the GW signal will exhibit a lower and broader peak at $k \sim \kstar/ e$, a narrower higher peak at $k=2/\sqrt{3} \kstar$ separated by a (shallow) dip at $k \sim \sqrt{2/3} \kstar$.

\subsubsection{The periodic structure of peaks in $\mathcal{P}(k)$ leads to a periodic modulation in $\Omega_{\textrm{GW}}(k)$}
\label{sec:Obs2}
One key signature of the GW signal for the two example power spectra (shown in blue and orange) in fig.~\ref{fig:OmegaGWofx_deltaetaperp} are the modulations visible on the principal peak. Here we will show how this feature is characteristic for a scalar power spectrum consisting of a comb of peaks in with a periodic structure.

To understand the genesis of the modulations it will be convenient to formally rewrite the power spectrum $\mathcal{P}(k)$ in \eqref{eq:P-analytic-xi-minus3} as a sum of the individual peaks:
\begin{align}
    \label{eq:P-as-sum}
    \mathcal{P}_\zeta(k) = \sum_{i=1}^{n_\textrm{p}} \mathcal{P}_{k_{\star i}} (k) \, ,
\end{align}
with $n_\textrm{p}$ the total number of individual peaks in $\mathcal{P}_\zeta(k)$ (as delimited by two adjacent zeros of $\mathcal{P}_\zeta(k)$) and where $\mathcal{P}_{k_{\star i}} (k)$ describes a single peak with maximum at $k_{\star i}$. 

As we have done with $\mathcal{P}_\textrm{env}$ when analysing the case of the overall shape of the GW signal, here it will be instructive to model the individual peaks $\mathcal{P}_{k_{\star i}}$ as $\delta$-distributions. Thus consider the following scalar power spectrum:
\begin{align}
\label{eq:P-as-delta-comb}
 \mathcal{P}_\zeta^{(\delta, \textrm{n}_\textrm{p})}(k) = \sum_{i=1}^{n_\textrm{p}} \mathcal{A}_i \, \delta (k-k_{\star i}) \, ,  
\end{align}
where we define $\mathcal{A}_i$ as the area (integral over $k$) under an individual peak $\mathcal{P}_{k_{\star i}}$. The GW energy density power spectrum $\Omega_{\textrm{GW}}$ for a comb of $\delta$-peaks has been computed in \cite{Cai:2019amo}, which can be done analytically. Inserting \eqref{eq:P-as-delta-comb} into \eqref{eq:OmegaGW-i} we find
\begin{align}
    \label{eq:Omega-from-delta-comb}
    \Omega_\textrm{GW}^{(\delta, \textrm{n}_\textrm{p})}(k) = c_g \Omega_{\textrm{r},0} \sum_{i,j=1}^{\textrm{n}_\textrm{p}} \frac{\mathcal{A}_i \mathcal{A}_j}{3k^2} \mathcal{T}_{\textrm{RD}} \Big(\tfrac{|k_{\star i}- k_{\star j}|}{\sqrt{3}k}, \tfrac{k_{\star i}+ k_{\star j}}{\sqrt{3}k} \Big) \, \Theta(k_{\star i} + k_{\star j}-k) \Theta(k - |k_{\star i} - k_{\star j}|) \, ,
\end{align}
with $\mathcal{T}_{\textrm{RD}}(d,s)$ defined in \eqref{eq:TRDds}.
This expression exhibits a series of resonance peaks at $k_{\textrm{max},ij}$ where $\Omega_\textrm{GW}^{(\delta, \textrm{n}_\textrm{p})}(k)$ diverges with \cite{Cai:2019amo}:\footnote{This is a $\log$-divergence which occurs whenever the denominator in the $\log$-term in \eqref{eq:TRDds} vanishes.}
\begin{align}
\label{eq:kmaxij-def}
k_{\textrm{max},ij} = \frac{1}{\sqrt{3}}(k_{\star i}+ k_{\star j}) \, , \quad \textrm{with} \quad k_{\textrm{max},ij} > |k_{\star i}- k_{\star j}| \, ,
\end{align}
where the restriction is enforced by the second $\Theta$-function in \eqref{eq:Omega-from-delta-comb}. 

If one replaces the $\delta$-peaks in \eqref{eq:P-as-delta-comb} by finite-width peaks, the divergences disappear. However, if the peaks in $\mathcal{P}_\zeta$ are sufficiently narrow, there will still be a visible local maximum in $\Omega_{\textrm{GW}}$ at the location $k_{\textrm{max},ij}$. This has been e.g.~observed for a scalar power spectrum consisting of a comb of narrow Gaussian peaks in \cite{Cai:2019amo}. Here this gives rise to the modulations in $\Omega_{\textrm{GW}}$ visible in fig.~\ref{fig:OmegaGWofx_deltaetaperp}. The individual peaks $\mathcal{P}_{k_{\star i}}$ comprising the power spectrum are sufficiently narrow to generate a series of peaks in $\Omega_{\textrm{GW}}$ located at $k \sim k_{\textrm{max},ij}$. We observe these peaks as the maxima of the modulations in $\Omega_{\textrm{GW}}$ as can be seen explicitly in fig.~\ref{fig:Omega-delta-etap-vlines}. There we plot $h^2 \Omega_{\textrm{GW}}$ for the example with $(\delta, \etaperp)=(\deA,\etaA)$ together with the $\delta$-peak result \eqref{eq:Omega-from-delta-comb}. One observes that the maxima of all visible modulations indeed closely track $k_{\textrm{max},ij}$ as asserted.\footnote{Note that there is a multitude of loci $k_{\textrm{max},ij}$ in fig.~\ref{fig:Omega-delta-etap-vlines} which do not correspond to visible peaks in $\Omega_{\textrm{GW}}$. This is particularly the case for the loci with the smallest and largest values of $k_{\textrm{max},ij}$. The reason is that any peak in $\Omega_{\textrm{GW}}$ there would have to be sourced by the constituent peaks $\mathcal{P}_{\kstar i}$ with the smallest and largest values of $k_{\star i}$. These peaks, however, have a small amplitude compared to the central peaks near $k \sim \kstar$ where the envelope $\mathcal{P}_\textrm{env}(k)$ is largest. As a result, the GW signal from the peaks $\mathcal{P}_{\kstar i}$ with the smallest and largest values of $k_{\star i}$ is suppressed compared to that of the central peaks and hence does not give rise to visible peaks in $\Omega_{\textrm{GW}}$.}

Last, we show that the periodicity of the peaks in $\mathcal{P}_\zeta(k)$ translates into a periodicity of the modulations in $\Omega_{\textrm{GW}}$. As observed in sec.~\ref{sec:enhancingP}, to a good approximation the positions $k_{\star i}$ of the constituent peaks $\mathcal{P}_{k_{\star i}}(k)$ can be treated as periodic in $k$:
\begin{align}
    k_{\star (i+1)}-k_{\star i} = \Delta k \, ,
\end{align}
with $\Delta k$ given in \eqref{eq:kperiod}. According to the above analysis this then implies that two neighbouring modulation maxima in $\Omega_{\textrm{GW}}$ are separated as
\begin{align}
    k_{\textrm{max},(i+1)j} - k_{\textrm{max},ij} =k_{\textrm{max},i(j+1)} - k_{\textrm{max},ij} = \frac{\Delta k}{\sqrt{3}} \, ,
\end{align}
i.e.~the modulations inherit the periodicity in $k$ as asserted.

The observations here described can then be summarised as follows: A scalar power spectrum consisting of a series of individual peaks will in turn give rise to a series of peaks in $\Omega_{\textrm{GW}}$ giving rise to a modulation in the GW signal. 
If the peaks in $\mathcal{P}_\zeta$ are periodic in $k$ with frequency $\omegalin$ the resulting modulation in $\Omega_{\textrm{GW}}$ will inherit this periodicity but with frequency $\omegagwlin=\sqrt{3}\,\omegalin$.\footnote{The precise numerical factor $\sqrt{3}$ ultimately comes from the fact that scalar-induced GWs are generated during the radiation era, with equation of state $w=1/3$. One can hence imagine that this numerical factor relating $\omegalin$ to $\omegagwlin$ can depend more precisely on the expansion history of the universe, see e.g.~\cite{Hajkarim:2019nbx,Domenech:2019quo,Domenech:2020kqm} about the implications of non-standard thermal histories on scalar-induced SGWBs.}

\vspace{0.1cm}

\noindent \textbf{A brief look at resonant features. --} We focus on sharp features in this paper, but it is interesting to try to generalise the reasoning above to the oscillatory patterns of resonant features, with an envelope of ${\cal P}_\zeta$ modulated by characteristic logarithmic oscillations in $\cos(\omegalog \log(k/k_{\textrm{ref}})+\phi)$. With the same notations and assumptions as above, one can apply Eqs.~\eqref{eq:Omega-from-delta-comb}-\eqref{eq:kmaxij-def}, now with the loci of the peaks $k_i$ that are equally spaced in $\log(k)$, i.e. $k_{\star i} \propto e^{2\pi i/\omegalog}$. If the frequency $\omegalog$ is smaller than a critical value $\omegalogc=2 \pi/ \log((\sqrt{3}+1)/(\sqrt{3}-1))\simeq 4.77$, one finds that different peaks $i\neq j$ are too widely spaced to ``interact'', i.e.~the condition $k_{\textrm{max},ij} > |k_{\star i}- k_{\star j}|$ in Eq.~\eqref{eq:kmaxij-def} is not satisfied. In that case, $\Omega_{\textrm{GW}}$ exhibits a series of resonance peaks at locations $k_{\textrm{max},ii} = \frac{2}{\sqrt{3}}k_{\star i}$, simply corresponding to the expected peaks in $\Omega_{\textrm{GW}}$ from each individual peak in ${\cal P}_\zeta$. These peaks are equally spaced in $\log(k)$, with the same frequency $\omegalog$ as the one characterising the curvature power spectrum. Hence, we have shown that for $\omegalog< \omegalogc$, the modulation in $\Omega_{\textrm{GW}}$ inherit the log-periodicity of ${\cal P}_\zeta$, with the same frequency, i.e. $\omegagwlog=\omegalog$.
If $\omegalogc \leq \omegalog < 2 \omegalogc$, nearest neighbouring peaks also interact, but not more distant ones, generating another series of intermediate peaks, of loci $k_{\textrm{max},i (i+1)}$ comprised between the ``principal peaks'' $k_{\textrm{max},ii}$, and with the same log-periodicity $\omegagwlog=\omegalog$. The pattern repeats itself, and for $n\, \omegalogc \leq \omegalog < (n+1)\, \omegalogc$, one can see that each peak interacts with its $n$ nearest neighbours, resulting again in a periodic structure in $\log(k)$ of same frequency $\omegalog$. 

For sufficiently large $\omegalog$ one can show that the GW spectrum will also exhibit a log-periodic structure with frequency $2 \omegalog$. To see this, consider for example the expected locations of the maxima in $\Omega_{\textrm{GW}}$ from interactions between the $(i+n)$-th and $(i+m)$-th peaks in the scalar power spectrum and compare those to the expected maximum due to the $i$-th peak. Using \eqref{eq:kmaxij-def} one finds
\begin{align}
\label{eq:kmaxij-inim-over-ii}
\frac{k_{\textrm{max},(i+n)(i+m)}}{k_{\textrm{max},ii}} = \cosh \bigg(\frac{\pi}{\omegalog} (n-m) \bigg) e^{\frac{\pi}{\omegalog} (n+m)} \, .
\end{align}
For sufficiently large $\omegalog$ and a scalar power spectrum with a sufficiently high number of peaks, one can then identify a range of values of $n$ and $m$, which satisfy
\begin{align}
    \frac{\pi (n-m)}{\omegalog} \ll 1 \lesssim \frac{\pi (n+m)}{\omegalog} \, .
\end{align}
For this range of values we can approximate the $\cosh$-term in \eqref{eq:kmaxij-inim-over-ii} by unity. The factor $\exp \big(\tfrac{\pi}{\omegalog} (n+m)\big)$ remains and corresponds to peaks spaced in $\log k$, but with frequency $2 \omegalog$.

In addition, larger values of $\omegalog$, leading to more interactions between distant peaks, are likely to diminish the magnitude of the resulting oscillation in $\Omega_\textrm{GW}$, as can be expected on general grounds. This is confirmed by preliminary numerical investigations. 

The findings described here indicate that the GW energy fraction, for some range of scales $k$, can be reproduced by the template 
\begin{align}
\label{eq:resonant_template}
\Omega_{\textrm{GW}}(k) = \overline{\Omega}_{\textrm{GW}}(k) \Big[1 &+ A_{1} \cos \big(\omega_\textrm{log} \log (k/k_\textrm{ref}) + \phi_{1} \big) \\
\nonumber & + A_{2} \cos \big(2 \omega_\textrm{log} \log (k/k_\textrm{ref}) + \phi_{2} \big) \Big] \, ,
\end{align}
where $k_\textrm{ref}$ is a reference scale that can be chosen freely and $\overline{\Omega}_{\textrm{GW}}(k)$ is a model-dependent envelope that is sufficiently smooth on length scales of the oscillations. A more detailed analysis of this will be the subject of a dedicated work \cite{resonant-SGWB}.

\subsubsection{The higher the number of visible modulations on the principal peak in $\Omega_{\textrm{GW}}$ the smaller their amplitude.}
\label{sec:Obs3}
We return to the plots of $\Omega_\textrm{GW}$ vs.~$\x=k/(\kf \etaperp)$ for the two examples with $(\delta, \etaperp)= (\deA,\etaA)$ and $(\deB,\etaB)$ in fig.~\ref{fig:OmegaGWofx_deltaetaperp}. As argued above, the overall shape of the principal peak (in $\x$-space) is the same for the two examples as they share the same value for $\deltaeta= \deeta$. We then make the following observations regarding the modulations on the principal peak: Firstly, the result for the model with $(\deB,\etaB)$ has more visible modulations than the GW signal for $(\deA,\etaA)$. Secondly, the modulations for the example with $(\deB,\etaB)$ appear to have a smaller amplitude than those for $(\deA,\etaA)$. The term amplitude here refers to the difference in value of $\Omega_{\textrm{GW}}$ compared to a smooth featureless principal peak, like the GW signal for the envelope power spectrum shown as the dashed green line in fig.~\ref{fig:OmegaGWofx_deltaetaperp}.

\begin{figure}[t]
\centering
\begin{overpic}[width=1.0\textwidth]{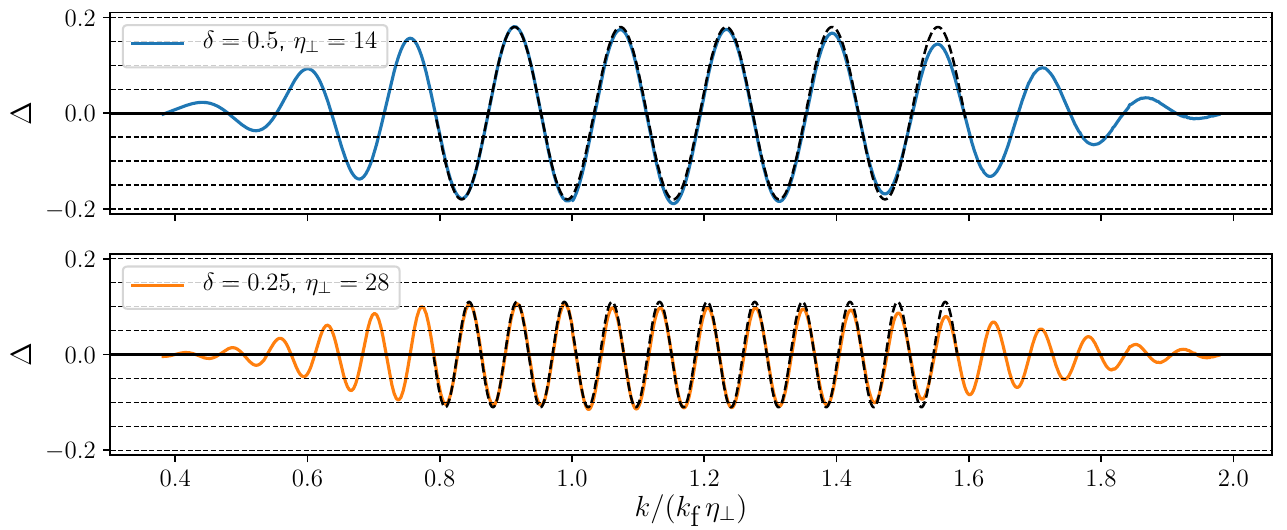}
\end{overpic}
\caption{\textit{Amplitude $\OmegaRatio(k)$ of modulations on the principal peak of the GW spectrum as defined in \eqref{eq:amplitude-of-modulations}. The blue plot in the upper panel is for model parameters $(\delta, \etaperp)=(\deA,\etaA)$ and the orange plot in the lower panel for $(\deB,\etaB)$. A fit of the simple template \eqref{eq:template-lin-maintext} to the central part of the GW signals is shown as the dashed black line.}}
\label{fig:Omega-amplitude-deltaetaperp}
\end{figure}

The first observation regarding the number of modulations can be understood as follows. As we have argued above, for the examples considered the modulations in $\Omega_{\textrm{GW}}$ are periodic in $k$ with period $\Delta k / \sqrt{3}$, where $\Delta k$ is the separation between the individual peaks in $\mathcal{P}_\zeta(k)$. In $\x$-space the corresponding period is $\Delta \x / \sqrt{3}$ which here is related to the model parameters as
\begin{align}
    \frac{\Delta \x}{\sqrt{3}} \approx \frac{\pi}{\sqrt{3}} \frac{e^{\delta / 2}}{\etaperp} \underset{\delta \ll 1}{\approx} \frac{\pi}{\sqrt{3} \, \etaperp} \, ,
\end{align}
where we have used the approximate expression \eqref{eq:xperiod} for $\Delta \x$. This implies that in $\x$-space the modulations are more closely bunched for larger values of $\etaperp$ and hence we expect to see more modulations in a fixed $\x$-interval. This is indeed what we observe in fig.~\ref{fig:OmegaGWofx_deltaetaperp} where we plot $\Omega_{\textrm{GW}}$ vs.~$\x=k / (\kf \etaperp)$. The upshot is that for models with fixed $\deltaeta$ (i.e.~fixed shape of the principal peak in $\x$-space) the number of visible modulations increases for larger values of $\etaperp$.   

We now turn to the amplitude of the modulations. We define the quantity
\begin{align}
\label{eq:amplitude-of-modulations}
\OmegaRatio(k) \equiv \frac{\Omega_{\textrm{GW}}(k)-\overline{\Omega}_{\textrm{GW}}(k)}{\overline{\Omega}_{\textrm{GW}}(k)} \, ,
\end{align}
which gives the magnitude of modulations of the GW spectrum over a smoothed featureless GW signal $\overline{\Omega}_{\textrm{GW}}$ as a fraction of that signal. Here we will use for $\overline{\Omega}_{\textrm{GW}}$ the GW fraction computed for the envelope power spectrum $\mathcal{P}_\textrm{env}$ shown as the dashed green line in fig.~\ref{fig:OmegaGWofx_deltaetaperp}. In fig.~\ref{fig:Omega-amplitude-deltaetaperp} we plot $\OmegaRatio(k)$ for the two example models with $(\delta, \etaperp)=(\deA,\etaA)$ and $(\deB,\etaB)$ over the range of $k$ of the principal peak. We observe that $\Delta(k)$ is sinusoidal throughout, with near-constant amplitude for $0.8 \lesssim k / (\kf \etaperp) \lesssim 1.6$, i.e.~over the central range of the principal peak. For the model with $(\delta, \etaperp)=(\deA,\etaA)$ this amplitude is $\Delta_\textrm{max} \approx 18 \%$ while for the model with $(\deB,\etaB)$ this is reduced to $\Delta_\textrm{max} \approx 11\%$.

This observation suggests that the essence of the GW spectrum across the principal peak is captured by the simple template
\begin{align}
\label{eq:template-lin-maintext}
\Omega_\textrm{GW}(k)=\overline{\Omega}_{\textrm{GW}}(k) \Big(1+A \cos(\omegagwlin \, k +\phi)  \Big) \, ,
\end{align}
with $A=\Delta_\textrm{max}$, 
$\omegagwlin = \sqrt{3}\, \omegalin$ and $\phi$ some constant phase. A fit of \eqref{eq:template-lin-maintext} is shown as the black dashed line in fig.~\ref{fig:Omega-amplitude-deltaetaperp}. To reproduce the curves shown there we had to reduce $\omegagwlin$ by $5 \%$ and $1\%$ compared to the value 
$\omegagwlin = \sqrt{3}\, \omegalin$
for the two models with $(\delta, \etaperp)=(\deA,\etaA)$ and $(\deB,\etaB)$, respectively. This accounts for the fact that the separation of the extrema on the principal peak is slightly shifted compared to the value $\Delta k / \sqrt{3}$ due to the change of $\overline{\Omega}_{\textrm{GW}}(k)$ across the principal peak.
A more precise template can be obtained by allowing for a mild $k$-dependence in $A$ and $\omegagwlin$ to capture the change in amplitude across the principal peak as well as any slight drift in $\omegagwlin$ away from the centre. 

The reduction in amplitude from $\Delta_\textrm{max} \approx 18 \%$ to $\Delta_\textrm{max} \approx 11 \%$ between the two models can be understood as follows. Note that here the amplitude is observed to be diminished when the number of modulations on the principal peak is higher. This suggests the explanation that, as the number of modulations is increased, the modulation peaks increasingly overlap with one another, leading to an `averaging-out' effect that reduces the amplitude. 

\begin{figure}[t]
\centering
\begin{overpic}[width=1.0\textwidth]{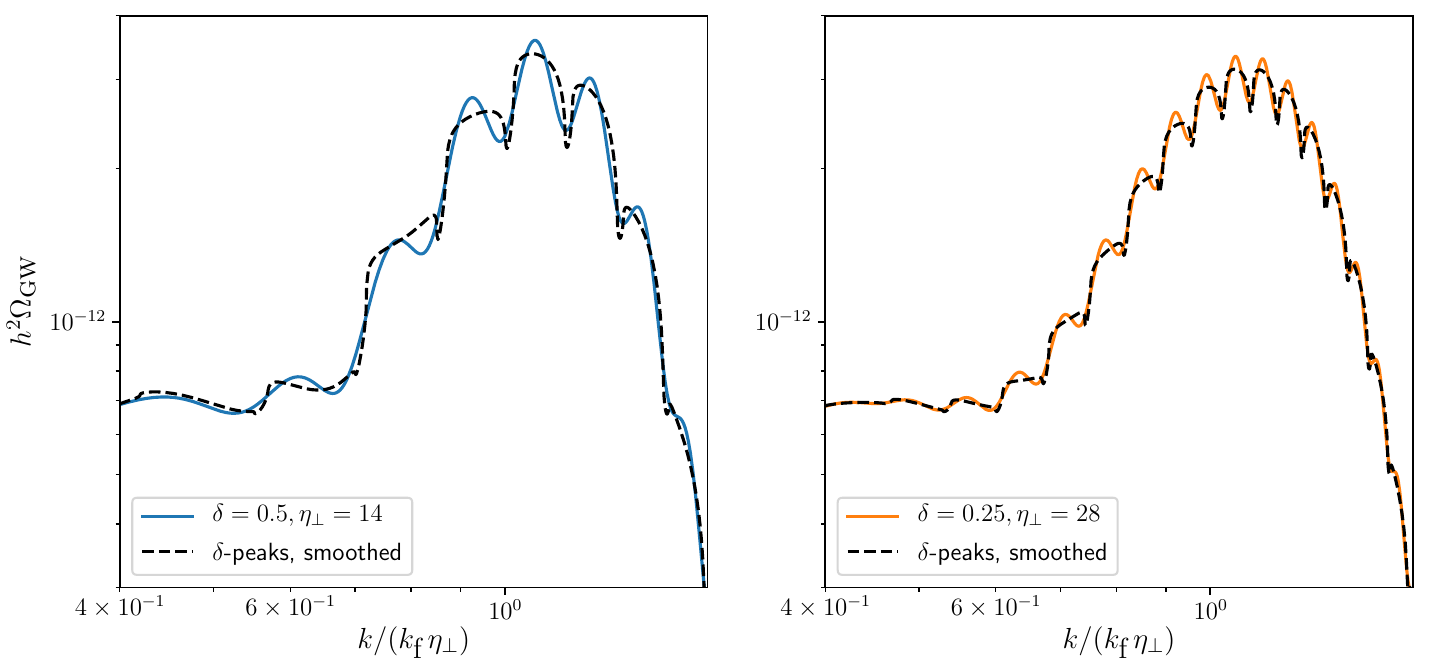}
\end{overpic}
\caption{\textit{$h^2 \Omega_{\textrm{GW}}(k)$ (blue/orange) and $h^2 \tilde{\Omega}_{\textrm{GW}}(k)$ (black, dashed) for the two parameter choices $(\delta,\etaperp)=(\deA,\etaA)$ and $(\deB,\etaB)$ shown in the LHS and RHS plots, respectively. The quantity  $\tilde{\Omega}_{\textrm{GW}}(k)$ is obtained by first computing the GW spectrum \eqref{eq:Omega-from-delta-comb} for the $\delta$-peak approximation \eqref{eq:P-as-delta-comb} of the scalar power spectrum before smoothing it over a scale $\Delta k /2$ as in \eqref{eq:Omega-ndelta-smoothed}.}}
\label{fig:Omega-delta-smoothed-deltaetaperp}
\end{figure}

\begin{figure}[t]
\centering
\begin{overpic}[width=1.0\textwidth]{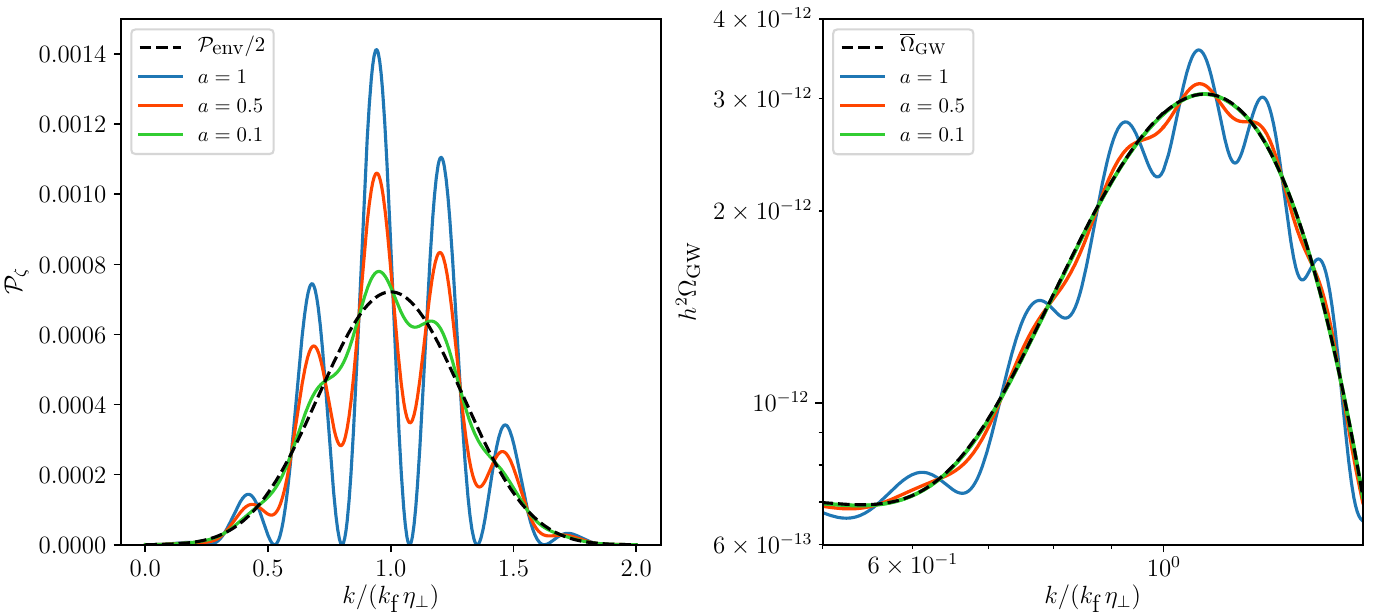}
\end{overpic}
\caption{\textit{LHS: Three example scalar power spectra $\mathcal{P}_\zeta(k)$ exhibiting a modulation over a common smooth curve $\mathcal{P}_\textrm{env}(k)/2$ with different values $a=1$, $0.5$ and $0.1$ (blue, red, green) for the amplitude of oscillation. The power spectra are given by \eqref{eq:P-analytic-xi-minus3} with $(\delta, \etaperp)= (\deA,\etaA)$, but a factor $a$ is inserted in front of both the $\cos$ and the $\sin$ terms in the numerator. RHS: $\Omega_\textrm{GW}(k)$ (restricted to the principal peak) for the three examples and  $\overline{\Omega}_\textrm{GW}(k)$ corresponding to the GW spectrum for $\mathcal{P}_\textrm{env}(k)/2$. The decrease in the amplitude of oscillations in $\mathcal{P}_\zeta(k)$ leads to a corresponding reduction of the amplitude of oscillations in $\Omega_\textrm{GW}(k)$. In particular, for the example with $a=0.1$ the modulation in $\Omega_\textrm{GW}(k)$ is practically imperceptible by eye.}}
\label{fig:PandOmegaGW_a1_0p5_0p1}
\end{figure}

This hypothesis can be tested explicitly. As a first step, we once more model the scalar power spectrum as a comb of $\delta$-peaks as in \eqref{eq:P-as-delta-comb}. The corresponding GW spectrum was computed in \eqref{eq:Omega-from-delta-comb} and plotted in fig.~\ref{fig:Omega-delta-etap-vlines} for the model with $(\delta, \etaperp)=(\deA,\etaA)$. As argued before, this result correctly predicts the location of the maxima of the modulations, but does not reproduce the amplitude of modulations as the resonance peaks diverge for the $\delta$-peak case. To account for the finite width of the individual peaks in the power spectrum $\mathcal{P}_\zeta(k)$ in \eqref{eq:P-analytic-xi-minus3} we can proceed as follows. The finite width of the peaks can be understood as entering the computation of the GW spectrum as a smoothing scale. This suggests that we can arrive at an estimate for $\Omega_{\textrm{GW}}$ for finite-width peaks by taking the $\delta$-peak result $\Omega_{\textrm{GW}}^{(\delta, n_{\textrm{p}})}$ and smoothing it over the width of an individual peak. For the power spectrum $\mathcal{P}_\zeta(k)$ in \eqref{eq:P-analytic-xi-minus3} the width of a constituent peak (defined as the full width at half-maximum) was found to be equal to half the separation between peaks, i.e.~$\Delta k/2$ with $\Delta k$ given in \eqref{eq:kperiod}. We then define the smoothed GW spectrum $\tilde{\Omega}_{\textrm{GW}}(k)$ as
\begin{align}   
\label{eq:Omega-ndelta-smoothed}
   \tilde{\Omega}_{\textrm{GW}}(k) = \frac{1}{\Delta k /2} \int_{k-\Delta k /4}^{k+\Delta k /4} \textrm{d}q \, \Omega_{\textrm{GW}}^{(\delta, n_{\textrm{p}})} (q) \, .
\end{align}
In fig.~\ref{fig:Omega-delta-smoothed-deltaetaperp} we plot $\tilde{\Omega}_{\textrm{GW}}(k)$ for the our two parameter choices $(\delta, \etaperp)=(\deA,\etaA)$ and $(\deB,\etaB)$ together with the full result $\Omega_{\textrm{GW}}(k)$, finding that the approximate expression $\tilde{\Omega}_{\textrm{GW}}(k)$ gives a good estimate of the amplitude of modulations. Most importantly, the reduction in amplitude from the model $(\deA,\etaA)$ to $(\deB,\etaB)$ is reproduced by the smoothed result $\tilde{\Omega}_{\textrm{GW}}(k)$ giving evidence that this observation is indeed caused by an `averaging-out' effect. As a result, while the oscillations in $\mathcal{P}_\zeta(k)$ are $\mathcal{O}(1)$ in the sense that the power spectrum drops all the way to zero between peaks, the resulting oscillations in $\Omega_{\textrm{GW}}(k)$ are only of the order $\sim \mathcal{O}(10 \%)$.   

\vspace{0.1cm}

\noindent \textbf{Decreasing the amplitude of oscillations in $\mathcal{P}_\zeta(k)$. --} Another property of the scalar power spectrum that will affect the amplitude of the modulations in $\Omega_\textrm{GW}$ is the amplitude of the oscillations in $\mathcal{P}_\zeta(k)$. So far we have assumed that the oscillations in $\mathcal{P}_\zeta(k)$ are $\mathcal{O}(1)$, as observed for the power spectrum \eqref{eq:P-analytic-xi-minus3} employed in our analysis. Here we will comment on how our findings are modified if the amplitude of oscillations in $\mathcal{P}_\zeta(k)$ is decreased, i.e.~for a scalar power spectrum that matches the template \eqref{eq:P-sharp-template} but with $a<1$.\footnote{Recall that for a sharp feature due to a period of copious particle production, a strong enhancement of the peak in $\mathcal{P}_\zeta(k)$ implies the presence of $\mathcal{O}(1)$ oscillations, see sec.~\ref{sec:generic}. Here, we relax this by hand to study the consequences.} For a smaller value of $a$ the scalar power spectrum resembles more and more its envelope, and hence the corresponding GW spectrum should increasingly resemble that of the envelope. The expectation thus is that our results from this section and sections \ref{sec:Obs1}, \ref{sec:Obs2} above are unchanged qualitatively, but that the amplitude of oscillations in $\Omega_\textrm{GW}$ is further reduced as $a$ is decreased. This is indeed what is observed in practice. In fig.~\ref{fig:PandOmegaGW_a1_0p5_0p1} we show $\mathcal{P}_\zeta(k)$ and the corresponding $\Omega_\textrm{GW}(k)$ (restricted to the principal peak) for three example power spectra exhibiting a modulation over a common smooth curve, but with a different amplitude of oscillations, respectively. The example with $a=1$ is again for the power spectrum \eqref{eq:P-analytic-xi-minus3} with parameter choices $(\delta, \etaperp)= (\deA,\etaA)$ employed throughout this and previous sections. The other two examples for $a=0.5$ and $a=0.1$ are for the power spectrum \eqref{eq:P-analytic-xi-minus3} with $(\delta, \etaperp)= (\deA,\etaA)$, but a factor $a$ is inserted in front of both the $\cos$ and the $\sin$ terms in the numerator. One can check that for the examples with $a=0.5$ and $a=0.1$ the corresponding modulation in $\Omega_\textrm{GW}$ is $\Delta_\textrm{max} \sim 4 \%$ and $\Delta_\textrm{max} \sim 0.2 \%$, respectively, while for $a=1$ we had $\Delta_\textrm{max} \sim 18 \%$. That is, the reduction of the amplitude of oscillations in $\Omega_\textrm{GW}$ is faster than proportional to the decrease in $a$.
 
\subsection{Observational prospects in gravitational wave observatories}
\label{sec:GWpredictions}
In this section we will discuss observational prospects for the scalar-induced SGWB from sharp turns during inflation in forthcoming GW observatories. To arrive at quantitative results we need to choose specific models and here we will once more focus on the GW signal produced by models with $\xi=-3$ and a turn with top-hat profile. The relevant scalar power spectrum is given in \eqref{eq:P-analytic-xi-minus3}.

As described in the previous section, the SGWB signal arising in this class of models exhibits a broad lower peak and a narrow higher principal peak with modulations, separated from one another by a shallow dip. We begin with some general observations regarding the relation between the model parameters $\kf$, $\delta$, $\etaperp$ and properties of the GW spectrum such as the amplitude and location (in frequency) of the principal peak as well as the period of modulations.

\begin{figure}[t]
\centering
\begin{subfigure}{.5\textwidth}
 \centering
   \begin{overpic}
[width=1.0\textwidth]{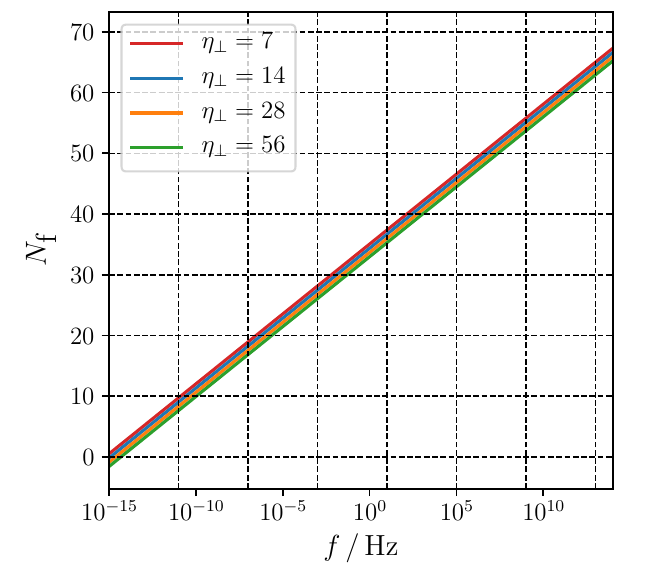}
\end{overpic}
\caption{\hphantom{A}}
\label{fig:Nf-of-fHz}
\end{subfigure}%
\begin{subfigure}{.5\textwidth}
 \centering
   \begin{overpic}
[width=1.0\textwidth]{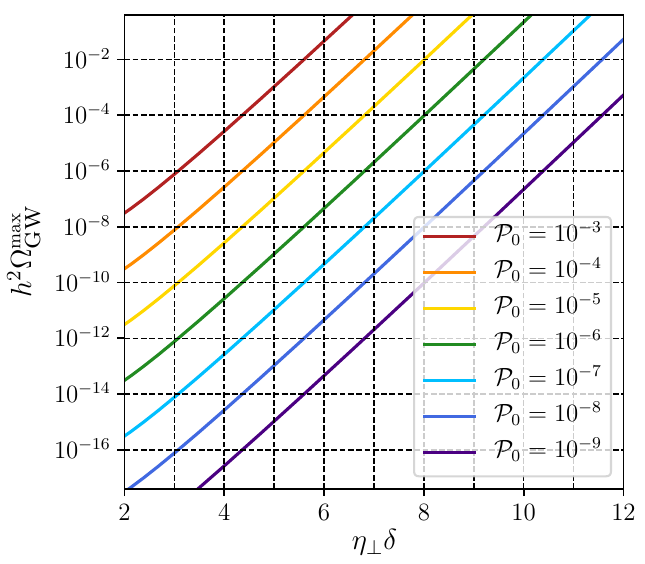}
\end{overpic}
\caption{\hphantom{A}}
\label{fig:Omegamax-of-deltaetaperp}
\end{subfigure}%
\caption{\textbf{(a):} \textit{$N_\textrm{f}$ vs.~$f_\textrm{peak}/$Hz in \eqref{eq:Nf-as-function-of-f} for $\etaperp=7, \, 14, \, 28, \, 56$.} \textbf{(b):} \textit{Estimate of the maximal amplitude $h^2 \Omega_{\textrm{GW}}^{\textrm{max}}$ in \eqref{eq:Omega-max-result-full} as a function of $\deltaeta$ for various values of $\Pzero$.}}
\label{fig:Nf-and-Omegamax}
\end{figure}

\vspace{0.1cm}

\noindent \textbf{Frequency of gravitational waves at the principal peak. --} In section \ref{sec:GWproperties} the location of the principal peak was determined as $k_{\textrm{peak}} \sim 2 / \sqrt{3} \kf \etaperp \sim \kf \etaperp$. To relate this to the corresponding frequency $f_\textrm{peak}$ we first re-express $\kf=k_\textsc{cmb} e^{N_\textrm{f}}$ where $k_\textsc{cmb}$ is scale of CMB modes and $N_\textrm{f}$ is normalised to denote the number of $e$-folds after the sourcing of the CMB modes. Then, using $k_\textsc{cmb} = 0.05 \, \textrm{Mpc}^{-1}$ and $f = 1.5 \cdot 10^{-15} k \, \textrm{Mpc} \, \textrm{Hz}$ we find the following relation between the model parameters $N_\textrm{f}$, $\etaperp$ and the observable quantity $f_\textrm{peak}$: 
\begin{align}
\label{eq:Nf-as-function-of-f}
N_\textrm{f} + \log(\etaperp) \approx \log \bigg( \frac{f_\textrm{peak}}{\textrm{Hz}}\bigg) + 37 \, .
\end{align}
Current and forthcoming experiments will be sensitive to different frequency bands spanning the range $10^{-10} \, \textrm{Hz} < f < 10^3 \, \textrm{Hz}$, corresponding to $14 \lesssim N_\textrm{f} + \log(\etaperp) \lesssim 44$, but proposals exist to extend this range to the MHz and GHz regime (see \cite{Aggarwal:2020olq} for a recent white paper). In fig.~\ref{fig:Nf-of-fHz} we plot $N_\textrm{f}$ as a function of $f_\textrm{peak}/ \textrm{Hz}$ for various values of $\etaperp$.\footnote{Recall that for a strong turn $\eta_\perp \gg 1$. At the same time, arbitrarily strong turns are not physically possible in an explicit model, and constraints from backreaction and perturbative control also bound $\etaperp$, see section \ref{sec:control}.} To give an example, we find that for $\etaperp=\etaA$ and $\etaB$ we require $N_\textrm{f} \sim 27-30$ and $27-29$ respectively for the principal peak to be in the interval $f \sim 10^{-3} \, \textrm{Hz}-10^{-2} \, \textrm{Hz}$, which is the frequency band where the LISA observatory is expected to achieve maximal sensitivity \cite{LISA-L3}. This is compatible with inflation lasting $50-60$ $e$-folds while at the same time allowing CMB modes to exit the horizon before the onset of the turn.   

\vspace{0.1cm}

\noindent \textbf{Amplitude of the gravitational wave spectrum at the principal peak. --} Another observable is the absolute value of $\Omega_{\textrm{GW}}$ at the principal peak. This can be related to the model parameters through \eqref{eq:Omega-max-result} which we reproduce here for convenience:
\begin{align}
\label{eq:Omega-max-result-full}
h^2 \Omega_{\textrm{GW}}^{\textrm{max}} \approx 1.4 \cdot 10^{-6} \times \frac{\Pzero^2 \, e^{4 \deltaeta}}{(\deltaeta-1)} \bigg[{\bigg( \frac{1}{2} \log \bigg(\frac{3}{2 (\deltaeta-1)} \bigg) +1 \bigg)}^2 +1+ \frac{\pi^2}{2} \bigg] \, ,
\end{align}
where we have set $1/\mathcal{N}=1/4$ and $c_g \Omega_{r,0} h^2 = 1.6 \cdot 10^{-5}$. Note that this only depends on the model parameters $\delta$ and $\etaperp$ through the combination $\deltaeta$. A plot of $\Omega_{\textrm{GW}}^{\textrm{max}}$ vs.~$\deltaeta$ for various choices of $\Pzero$ is shown in fig.~\ref{fig:Omegamax-of-deltaetaperp}. 

From \eqref{eq:Omega-max-result-full} we can estimate the values of $\deltaeta$ required for our GW signal to be visible in various GW observatories for any given value of the `baseline' power spectrum $\Pzero$. The detection thresholds for a selected choice of experiments in their most sensitive frequency range are $h^2\Omega_{\textrm{GW}} \gtrsim 10^{-17}$ (BBO),
$h^2\Omega_{\textrm{GW}} \gtrsim 10^{-16}$ (DECIGO), $h^2\Omega_{\textrm{GW}} \gtrsim 10^{-13}$ (LISA, CE) and $h^2\Omega_{\textrm{GW}} \gtrsim 10^{-12}$ (ET), see e.g.~fig.~\ref{fig:Omega-of-f-sensitivity}.\footnote{The detection thresholds quoted here are chosen as the lowest values on the power-law-integrated sensitivity curves for $h^2 \Omega_{\textrm{GW}}$ for the various experiments. The notion of power-law-integrated sensitivity was defined in \cite{Thrane:2013oya} and the numbers quoted here are from \cite{Schmitz:2020syl}.} Let us first consider the case $\Pzero=\PCMB=2.4 \cdot 10^{-9}$, i.e.~we assume that there is no additional amplification of $\Pzero$ compared to $\PCMB$. From \eqref{eq:Omega-max-result-full} it then follows that for detection by the various experiments we require $\deltaeta \gtrsim 3.5$ (BBO), $\deltaeta \gtrsim 4$ (DECIGO), $\deltaeta \gtrsim 6$ (LISA, CE) and $\deltaeta \gtrsim 6.5$ (ET). We observe that for a turn with top-hat profile the angle $\Angle$ swept during the turn is given by $\Angle = \deltaeta$. Thus, to be visible in e.g.~LISA one requires $\Angle \gtrsim 2 \pi$. While this does not seem to be attainable in two-field models of inflation with a flat field space, this can be achieved if the field space is curved or, possibly, by invoking more than two fields. For instance, in negatively curved field spaces, trajectories with a strongly non-geodesic motion ($\eta_\perp \gg 1$) are commonplace, see references in footnote 23. A brief period of such motion can naturally give $\Angle=\eta_\perp \delta \gtrsim 2 \pi$. 

At this stage it is important to recall that the parameters $\delta$ and $\etaperp$ are constrained by considerations of backreaction and perturbative control. The condition for avoiding excessive backreaction on the inflationary background is given in \eqref{backreaction}.\footnote{There are further constraints for backreaction not to violate the `slow-roll' conditions $(\epsilon, \eta) \ll 1$. 
Anticipating that in an explicit model of inflation, a turn in the inflationary trajectory may coincide with a violation of slow-roll, we do not consider these conditions here any further.} A tentative analysis further suggests that perturbative control can be maintained if the condition \eqref{perturbativity} is satisfied. To examine what these constraints imply for the detectability of the GW signal in various observatories, we insert $\Pzero$ and the detection threshold values for $\deltaeta$ into \eqref{backreaction}-\eqref{perturbativity} and check to what extent these conditions can be satisfied while at the same time ensuring $\etaperp \gg \deltaeta$. For the values listed in the previous paragraph one finds that the condition \eqref{backreaction} for avoiding backreaction leaves sufficient parameter space for models that can be detected in BBO ($\etaperp \lesssim 98$), DECIGO ($\etaperp \lesssim 76$), LISA and CE ($\etaperp \lesssim 28$) and ET ($\etaperp \lesssim 22$). The condition \eqref{perturbativity} is  more stringent and implies that signals detectable in LISA and CE appear to be at the edge of theoretical control ($\etaperp \lesssim 7$) whereas signals visible in ET appear to be excluded ($\etaperp \lesssim 5.5$). Still, detection in BBO ($\etaperp \lesssim 31$) and DECIGO ($\etaperp \lesssim 24$) is not threatened. 

The above limits on $h^2 \Omega_{\textrm{GW}}^{\textrm{max}}$ and hence detectability posed by the backreaction and perturbativity constraints can be somewhat relaxed if $\Pzero$ is enhanced compared to $\PCMB$ independently of any amplification of scalar fluctuations due to the sharp turn. To give one example, if $\Pzero = 10^{-7}$, a signal that is detectable in LISA ($h^2\Omega_{\textrm{GW}} \gtrsim 10^{-13}$) can be achieved for $\deltaeta \gtrsim 4$. The constraints \eqref{backreaction}-\eqref{perturbativity} then require $\etaperp \lesssim 8$, which now appears to be marginally within the regime of control. However, increasing $\Pzero$ can only boost the attainable value of $h^2 \Omega_{\textrm{GW}}^{\textrm{max}}$ to a certain extent. For example, we find that the recent result by NANOGrav \cite{Arzoumanian:2020vkk}, which is consistent with a GW energy density power spectrum with $h^2 \Omega_{\textrm{GW}}^{\textrm{max}} \sim 10^{-9}$, remains outside of what can be achieved by strong and sharp turns: There is no region of parameter space that reproduces the required amplitude of the GW spectrum while at the same time satisfying \eqref{backreaction}-\eqref{perturbativity}. 

\vspace{0.1cm}

\noindent \textbf{Period and amplitude of modulation on the principal peak. --}
A characteristic feature of the GW spectrum from sharp turns are $\mathcal{O}(10 \%)$-level modulations on the principal peak. These modulations have an effectively constant period in $k$ given by $\Delta k \approx \pi e^{\delta/ 2} \kf / \sqrt{3}$.\footnote{This is a good approximation of the period of oscillation for $\delta \ll 1$. For more intermediate values $\delta <1$ a more accurate expression is given in \eqref{eq:period-near-max-1st}.} In frequency-space the peaks will appear at intervals $\Delta f$ with
\begin{align}
    \label{eq:Deltafoverfpeak}
    \frac{\Delta f}{f_\textrm{peak}} \approx \frac{\pi}{\sqrt{3}} \frac{e^{\delta/2}}{\etaperp} \, .
\end{align}
That is, the observable ratio $\Delta f / f_{\textrm{peak}}$ is determined by $\delta$ and $\etaperp$, but in a different combination than that which enters the amplitude $\Omega_{\textrm{GW}}^{\textrm{max}}$. Thus, a measurement of both $\Delta f / f_{\textrm{peak}}$ and $\Omega_{\textrm{GW}}^{\textrm{max}}$ in principle allows one to determine $\delta$ and $\etaperp$ separately. 

Another observable is the amplitude $\Delta_\textrm{max}$ of the modulations about a smoothed featureless signal. As described in sec.~\ref{sec:GWproperties}, the amplitude $\Delta_\textrm{max}$ decreases as $\etaperp$ is increased for fixed $\deltaeta$ (see fig.~\ref{fig:Omega-amplitude-deltaetaperp}). Thus, a measurement of the amplitude of the modulations provides another independent avenue for determining $\etaperp$ for a given value of $\deltaeta$.\footnote{It appears difficult to extract a simple analytical expression that relates $\delta$, $\etaperp$ and the amplitude of modulations. In absence of such a formula the determination of $\etaperp$ can be performed by matching the observed amplitude to numerically computed templates like those in fig.~\ref{fig:Omega-amplitude-deltaetaperp}.} 
This hence provides a cross-check for the values of $\delta$ and $\etaperp$ extracted from a measurement of $\Delta f / f_{\textrm{peak}}$ and $\Omega_{\textrm{GW}}^{\textrm{max}}$. It can be used to distinguish between different inflation models with sharp turns, where the dependence of $\Delta f / f_{\textrm{peak}}$, $\Omega_{\textrm{GW}}^{\textrm{max}}$ and $\Delta_\textrm{max}$ on $\delta$ and $\etaperp$ will differ between the various models.\\

We now return to fig.~\ref{fig:Omega-of-f-sensitivity} where we plotted $h^2 \Omega_{\textrm{GW}}$ vs.~$f / \textrm{Hz}$ for two example models with  $(\delta, \etaperp)=(\deA,\etaA)$ and $(0.5,10)$ together with sensitivity curves for a selected choice of GW observatories. 
The sensitivity curves shown here have been computed in \cite{Schmitz:2020syl} and correspond to the power-law-integrated sensitivity (PLIS) curves defined in \cite{Thrane:2013oya}.\footnote{These were computed in \cite{Schmitz:2020syl} for a threshold signal-to-noise-ratio $\textrm{SNR}_\textrm{threshold}=1$ and a total observation time of $T_\textrm{obs}=1$ year for all future interferometers and $T_\textrm{obs}=20$ years for all future pulsar timing arrays.} We assumed a top-hat profile for the turn and set $\Pzero=\PCMB = 2.4 \cdot 10^{-9}$ and $\xi=-3$. Note that the parameter choice $(\delta, \etaperp)=(\deA,\etaA)$ satisfies the condition \eqref{backreaction} for avoiding excessive backreaction on the inflationary background, but violates the tentative bound \eqref{perturbativity} for retaining perturbative control, while both bounds are satisfied by the model with $(\delta, \etaperp)=(0.5,10)$. The various copies of the GW spectrum in fig.~\ref{fig:Omega-of-f-sensitivity} are for different choices of the parameter $N_\textrm{f}$ which controls the frequency range of our GW signal. For the parameter choice $(\delta, \etaperp)=(\deA,\etaA)$ we show the GW spectrum for $N_\textrm{f}=15, \, 28.5, \, 33, \, 37$ while for $(\delta, \etaperp)=(0.5,10)$ we picked $N_\textrm{f}=14.75, \, 32.5$. The choice was made so that the principal peak of the GW signal appears in the frequency range of maximal sensitivity of the various GW observatories. The GW signal for the model with $(\delta, \etaperp)=(\deA,\etaA)$ can be potentially detected with IPTA, SKA, LISA, BBO, DECIGO, ET and CE, while the model with $(\delta, \etaperp)=(0.5,10)$ is potentially detectable in BBO and DECIGO and at the edge of detectability in SKA. Most notably, both the GW spectra for $(\delta, \etaperp)=(\deA,\etaA)$ and $(0.5,10)$ exhibit modulations on the principal peak, whose amplitudes (i.e.~the maximal value of \eqref{eq:amplitude-of-modulations}) are $18 \%$ and $22 \%$ of the total signal, respectively.

\begin{figure}[t]
\centering
\begin{overpic}[width=0.8\textwidth]{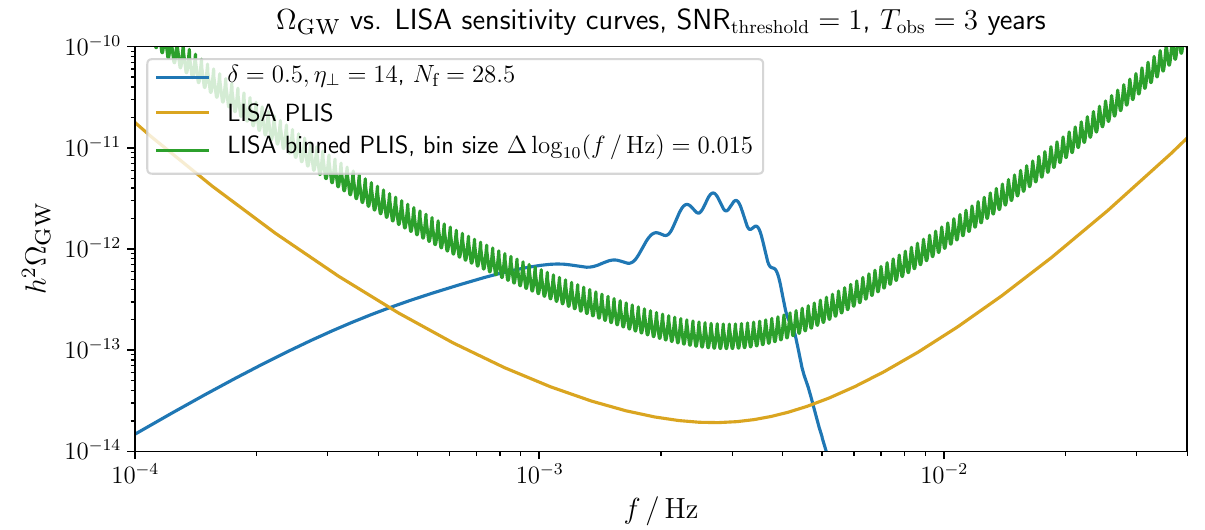}
\end{overpic}
\caption{\textit{$\Omega_\textrm{GW}(f)$ for the example model with parameters $(\delta, \etaperp)=(\deA,\etaA)$ and $\Nf=28.5$, together with the PLIS curve of \cite{Thrane:2013oya} and the binned PLIS curve of \cite{Caprini:2019pxz} for LISA. The PLIS curves are computed for a threshold signal-to-noise-ratio $\textrm{SNR}_\textrm{threshold}=1$ and a total observation time of $T_\textrm{obs}=3$ years. For the binned PLIS the size of each bin is chosen as $\Delta \log_{10} (f \, / \, \textrm{Hz}) = 0.015$.}}
\label{fig:LISAPLIS}
\end{figure}

However, the PLIS curves shown in figure \ref{fig:Omega-of-f-sensitivity} assume that the GW spectrum is given by a single power law over the whole frequency range, and hence they can only give a rough indication whether our signal can be reconstructed by the various experiments. In particular, comparing the signal to the PLIS curve is not sufficient to determine whether the oscillations in $\Omega_\textrm{GW}$ can be experimentally resolved. A better criterion for this purpose is given by comparison to the `binned PLIS' \cite{Caprini:2019pxz,Flauger:2020qyi}. The idea behind this concept is to split the frequency range into a number of bins and compute the PLIS for each bin individually. If the GW spectrum is such that it can be modelled sufficiently accurately by a power law in each bin, reconstruction of the signal is deemed possible if the GW spectrum lies above the binned PLIS curve. To give an example, in fig.~\ref{fig:LISAPLIS} we show $\Omega_\textrm{GW}$ computed for the model $(\delta, \etaperp)=(\deA,\etaA)$ with $\Nf=28.5$ together with the LISA PLIS and the binned PLIS. To successfully reconstruct the oscillation, we will need several bins over a single period, as we need multiple power laws to model the signal. Here we choose bins with width $\Delta \log_{10} (f / \textrm{Hz})=0.015$, which for the given example results in $\approx 4$ bins per period of oscillation on the principal peak of $\Omega_\textrm{GW}$. Using the binned PLIS reduces the LISA sensitivity by about one order of magnitude compared to the PLIS for the example at hand. However, this is not a severe reduction given the significant increase of frequency bins. Most importantly, the principal peak of $\Omega_\textrm{GW}$ with its modulations for the model with $(\delta, \etaperp)=(\deA,\etaA)$ lies comfortably above the binned PLIS curve, suggesting that a reconstruction of the signal including the oscillations is not unrealistic. For another recent use of the binned PLIS for reconstructing GW spectra with oscillations from features during inflation, see \cite{Braglia:2020taf}.

What remains to be understood is how large the amplitude of oscillations needs to be to be visible in a GW observatory. As has been analysed in \cite{Caprini:2019pxz} a `wiggly' GW signal can be successfully reconstructed by LISA if the modulations are sufficiently pronounced and the overall amplitude of the GW signal is sufficiently high. However, there the amplitude is $\Delta_\textrm{max} \sim \mathcal{O}(1)$ whereas our models at best produce $\Delta_\textrm{max} \sim 0.2$.\footnote{Also, the oscillations of the signals studied in \cite{Caprini:2019pxz} correspond to a periodic modulation in $\log(k)$.} This is inherent to post-inflationary scalar-induced GWs, where $10 \%$-level modulations in $\Omega_\textrm{GW}$ are a direct consequence of $100 \%$-level oscillations in $\mathcal{P}_\zeta$, see sec.~\ref{sec:GWproperties}. As described in this work, the presence of oscillations in the primordial contribution to the SGWB gives a unique opportunity for testing inflation on small scales with gravitational waves. We thus believe it is important to further study the detectability of modulations in $\Omega_\textrm{GW}$.

\section{Discussion and outlook}
\label{sec:discussion}

Copious particle production due to a sharp feature during inflation is associated with an enhanced scalar power spectrum $\mathcal{P}_\zeta(k)$ with $\mathcal{O}(1)$-oscillations that are periodic in $k$. These oscillations are inherited by the scalar-induced SGWB sourced when the scalar fluctuations re-enter the horizon after inflation. In particular, the resulting GW energy density power spectrum $\Omega_\textrm{GW}(k)$ exhibits a peak modulated by $\mathcal{O}(10 \%)$-oscillations, also periodic in $k$. The fact that $\mathcal{O}(1)$-oscillations in $\mathcal{P}_\zeta$ result in $\mathcal{O}(10 \%)$-modulations in $\Omega_\textrm{GW}(k)$ can be explained by an `averaging-out' process that is inherent to the mechanism by which scalar-fluctuations are processed into tensor fluctuations. 
The frequency $\omegagwlin$ of $\Omega_\textrm{GW}(k)$ is related to the frequency in the primordial power spectrum $\omegalin$ by the simple relation $\omegagwlin=\sqrt{3}\, \omegalin$, where the subscript indicates that the oscillations are linearly spaced in $k$. The amplitude and number of visible dulations in $\Omega_\textrm{GW}$ depend on how rapid the oscillations in $\mathcal{P}_\zeta$ are compared to the change of the envelope of $\mathcal{P}_\zeta$. If the oscillations are fast compared to a change in the envelope there will be more visible modulations in $\Omega_\textrm{GW}$, but their amplitude will be suppressed due to the `averaging-out' effect. In contrast, if there are only a few pronounced oscillations in $\mathcal{P}_\zeta$ for a given change in the envelope, there will be fewer modulations in $\Omega_\textrm{GW}$, but their amplitude will be more pronounced. As a result, if the oscillations in $\mathcal{P}_\zeta$ are too rapid, the modulations in $\Omega_\textrm{GW}$ will become so suppressed as to become effectively invisible.

An explicit realisation of the above can be obtained in models of multifield inflation, and we have used this for explicit calculations in this work. In these models a strong and sharp turn in the inflationary trajectory behaves like a sudden feature, leading to subsequent particle production resulting in an enhanced scalar power spectrum with $\mathcal{O}(1)$-oscillations \cite{Palma:2020ejf,Fumagalli:2020adf}. The strength of the turn is quantified by the parameter $\etaperp(N)$, whose magnitude describes the departure of the inflationary trajectory from a geodesic as a function of $e$-folds. To create a sudden feature with the desired effect we require $\etaperp \gg 1$ for at most one $e$-fold, i.e.~a strong and sharp turn.

The enhancement factor of the scalar power spectrum and hence the amplitude of $\Omega_\textrm{GW}$ are exponentially sensitive to the angle $\Angle$ swept during the turn. For example, to produce a GW signal with $h^2 \Omega_\textrm{GW} \gtrsim 10^{-13}$ as to be visible in e.g.~LISA one requires $\Angle \gtrsim 2 \pi$. While this appears to be unattainable in two-field models of inflation with a flat field space, this may be achievable if the field space is curved or by invoking more than two fields.\footnote{Strongly non-geodesic motions have been under scrutiny, notably as they allow to sustain inflation on otherwise too steep potentials, and most examples have been worked out in highly curved field spaces (see e.g.~\cite{Cremonini:2010ua,Renaux-Petel:2015mga,Brown:2017osf,Mizuno:2017idt,Christodoulidis:2018qdw,Garcia-Saenz:2018ifx,Achucarro:2018vey,Bjorkmo:2019aev,Fumagalli:2019noh,Bjorkmo:2019fls,Christodoulidis:2019jsx,Aragam:2019khr,Bravo:2019xdo,Garcia-Saenz:2019njm,Chakraborty:2019dfh,Aragam:2020uqi}). Here, the strong bending is only transient, and our models are phenomenological, but it is interesting to note that a recent analysis \cite{Calderon-Infante:2020dhm} indicates that the swampland distance conjecture ties the maximally allowed degree of bending to the curvature of field space, allowing larger values of $\etaperp$ for larger field space curvature.}

To arrive at results that are independent of the precise microscopic realisation of the turn we specified inflation models by proposing profiles for $H(N)$ and $\etaperp(N)$ without referring to a Lagrangian. To focus on the effects caused by a feature in $\etaperp(N)$ we assumed $H(N)$ to be featureless, corresponding to a smooth background evolution. We then computed the scalar power spectrum for a set of simple turn profiles $\etaperp(N)$: For $\etaperp(N)$ with a top-hat profile this can be done analytically, while for a Gaussian profile we resorted to numerical analysis, finding qualitatively similar results.

To give an example, a turn with top-hat profile with amplitude $\etaperp=10$ and a duration $\delta \equiv \delta N =\deA$ gives rise to an amplification of the scalar power spectrum by a factor $\exp(2 \deltaeta) \sim 2 \cdot 10^4$ compared to the value $\PCMB = 2.4 \cdot 10^{-9}$ at CMB scales. This in turn gives rise to a GW energy density power spectrum (see fig.~\ref{fig:Omega-of-f-sensitivity}) that peaks at $h^2 \Omega_\textrm{GW} \sim 10^{-15}$ and is thus potentially detectable with BBO or DECIGO. To fall into the frequency band $0.1 \, \textrm{Hz} < f < 1 \, \textrm{Hz}$ where BBO and DECIGO are the most sensitive the turn has to occur some $32-35$ $e$-folds after the sourcing of CMB modes. Most notably, the GW energy density power spectrum $h^2 \Omega_\textrm{GW}$ exhibits a series of modulations whose amplitude is $\sim 22 \%$ of the total signal. Similarly, with different set of parameters, a SGWB with an oscillatory frequency profile can be detectable by LISA, albeit in a regime where theoretical control is less established.\\

There are several open questions and avenues for future work. One important set of questions concerns the embedding of our amplification mechanism in explicit models of inflation, which may result in additional features relevant for phenomenology:

\vspace{0.1cm}

\noindent \textbf{Violations of slow-roll or phases of ultra-slow-roll.--} To focus on the effect from a sharp turn we have so far assumed a smooth background evolution with a featureless $H(N)$. A logical generalisation then is to also allow for a non-trivial profile for $H(N)$ and study the joint effect of sharp turns and departures from a smooth expansion rate on $\mathcal{P}_\zeta$ and $\Omega_{\textrm{GW}}$. While CMB data severely restricts $H(N)$ on large scales, there is more freedom at small scales, allowing e.g.~for slow-roll violating phases or periods of ultra-slow-roll inflation  \cite{Tsamis:2003px,Kinney:2005vj,Namjoo:2012aa},  the latter of which have been shown to also lead to an amplification of scalar fluctuations (see, e.g.~\cite{Garcia-Bellido:2017mdw,Germani:2017bcs}). Explicit models would be useful in this context to study to what extent features in $\etaperp(N)$ and $H(N)$ occur together. For example, in the recent two-field models \cite{Braglia:2020eai, McDonough:2020gmn} exhibiting an inflationary trajectory with a turn, the turn always coincides with a disturbance in $H(N)$, as the inflaton slows down and speeds up around the time of the turn. Violations of slow-roll at the turn are also interesting as they may relax the bounds from backreaction. As described in sec.~\ref{sec:control}, backreaction of the enhanced scalar fluctuation is most problematic for maintaining slow-roll, but these bounds become void if the slow-roll conditions are not satisfied in the first place during the amplification of the scalar fluctuations. 

\vspace{0.1cm}

\noindent \textbf{A series of turns.--} In this work we have assumed a single sharp turn during inflation for simplicity, which may not be generic in an explicit model. One may imagine that in an explicit model, after a sharp turn deflecting the inflaton from its initial trajectory, the inflaton oscillates before settling down again in a stable trajectory. These oscillations can be understood as a series of turns with alternating sign for $\etaperp$, and can generate a resonant amplifications of fluctuations. It would be interesting to examine how such a series of turns, giving rise to the classical clock signal when the bending is moderate \cite{Chen:2011zf,Chen:2014joa}, affects $\mathcal{P}_\zeta$ when the bending is strong, and to study whether this leads to observable features in $\Omega_{\textrm{GW}}$.

\vspace{0.1cm}

\noindent \textbf{Effects of reduced sound speeds.--} We have considered a sharp strong turn of the inflationary trajectory in nonlinear sigma models, in which curvature and entropic perturbations otherwise propagate with the speed of light. More generally, in the effective field theory of inflationary fluctuations (see e.g.~\cite{Senatore:2010wk,Noumi:2012vr} in a multifield context), each type of perturbations can propagate with reduced sound speeds, and it would be interesting to generalise our findings to such setups. 

\vspace{0.1cm}

There are additional open questions leading to further promising investigations:

\vspace{0.1cm}

\noindent \textbf{Considerations of backreaction and perturbative control.--} An important set of constraints on the presented mechanism derives from the requirements of avoiding excessive backreaction and maintaining perturbative control. In sec.~\ref{sec:control} we performed a tentative analysis of these issues, finding the constraints \eqref{backreaction-result}--\eqref{Ln} respectively, with the perturbativity bound \eqref{Ln} the more constraining of the two. Applied to our model, \eqref{backreaction-result}--\eqref{Ln} effectively bound the maximally attainable amplification due to short turns of the scalar power spectrum compared to the single-field power spectrum $\Pzero$, which in turn constrains the amplitude and hence detectability of the corresponding GW signal. Given the importance of these bounds, and in particular \eqref{Ln}, for the phenomenology of the setup, we find it imperative to go beyond the tentative analysis leading to \eqref{Ln} by e.g.~computing the one-loop contribution to the power spectrum. We expect that such further studies dedicated to rigorously ensuring perturbative control, taking into account the backreaction of the particles produced during the turn, as well as going beyond the top-hat profile for the turn, may reduce the allowed parameter space, as well as give interesting twists to the phenomenology described in this paper.

\vspace{0.1cm}

\noindent \textbf{Production of primordial black holes.--}
The huge enhancement of scalar fluctuations due to a sharp or strong turn during inflation as a source for primordial black holes (PBHs) has been proposed in \cite{Palma:2020ejf,Fumagalli:2020adf}. Hence, the mechanism presented here can be jointly confronted with experiment through its predictions for the SGWB and its mass spectrum of PBHs. One consequence is that the existing bounds on the PBH mass spectrum (see e.g.~\cite{Carr:2020gox} for a recent review) will give constraints on the parameter space of the mechanism.

The PBH mass spectrum is highly sensitive to the tail of the probability density function of scalar fluctuations, as PBHs form from rare large fluctuations (see e.g.~\cite{Byrnes:2012yx}). If scalar fluctuations are Gaussian, PBHs making up a significant fraction of DM requires a scalar power spectrum $\mathcal{P}_\zeta \sim 10^{-2}$ so that the PDF has a sufficiently broad tail. However, if the PDF departs from a Gaussian there is a priori no direct link between the enhancement of the power spectrum and the PBH abundance. In this case even a model with a more moderately enhanced power spectrum, which may hence be less threatened by backreaction or perturbativity issues, could in principle lead to a significant amount of PBHs. 

This is of relevance here as inflation models with periods of strongly non-geodesic motion, i.e.~strong turns, have been shown to depart from Gaussian statistics and exhibit a characteristic pattern of (flattened-type) non-Gaussianity \cite{Garcia-Saenz:2018vqf, Fumagalli:2019noh, Ferreira:2020qkf}. The precise consequence of this for the tail of the PDF is not known and hence robust predictions for the PBH mass spectrum in inflation models with strong or/and sharp turns are currently not available. Hence, at this stage one cannot exclude the possibility that inflation models with turns exhibit a non-negligible PBH abundance even for $\mathcal{P}_\zeta < 10^{-2}$. Another important question is to what extent the oscillations in the power spectrum due to a sharp turn lead to features in the PBH abundance.\footnote{An initial investigation of the PBH abundance (assuming Gaussian statistics for simplicity) in an inflation model with a sharp turn indicates that the oscillations in the scalar power spectrum do not lead to an oscillatory feature in the PBH mass spectrum \cite{Fumagalli:2020adf}.} All of this motivates further theoretical investigations directed towards computing the PBH abundance in these models.

\vspace{0.1cm}

\noindent \textbf{Competition with inflationary-period GWs.--} Mechanisms during inflation that generate a large enhancement of curvature fluctuations may also lead to a substantial generation of gravitational waves \textit{during} inflation, see e.g.~\cite{Cook:2011hg,Garcia-Bellido:2017aan,Cai:2019jah,Ozsoy:2020ccy,Ozsoy:2020kat} for studies in different mechanisms. This primordial SGWB may dominate over the one produced at horizon re-entry of the enhanced scalar fluctuations, and is correlated with it in general.\footnote{In addition to the radiation-generated scalar-induced contribution to $\Omega_{\textrm{GW}}$, this leads to a purely primordial contribution, and to a mixed primordial-induced one, see e.g.~\cite{Garcia-Bellido:2017aan}.} We focused on the latter in this paper, and on the model-independent characterisation of it when the curvature perturbation exhibits properties of sharp features. Still, it is interesting and important to assess the relative importance of the inflationary-period GWs and of the post-inflationary ones in the mechanism studied here leading to copious particle production. Given the scope of this paper, we will report on this with due details in a dedicated publication \cite{primordial-GWs}, but here are the main characteristics: $\Omegainf(k)$ has a localised enhancement on scales $k \sim \kstar$ together with oscillatory features, like $\Omegarad$ studied here in detail, although details differ. Analytical estimates, backed up with numerical computations, show that around the relevant scale $\kstar$, the ratio $\Omegainf/\Omegarad \simeq 10^{-2} \epsilon^2 \etaperp^4$ (with the exponent $4$ very mildly depending on $\deltaeta$ quantifying the efficiency of the particle production). The suppression by $\epsilon^2$ is generic and is also present without any amplification of scalar fluctuations. The amplification by $\etaperp^4$ is tied to the fact that the burst of particle production sourcing GWs arises on sub-Hubble scales for the relevant modes: the larger $\etaperp$, the deeper inside the Hubble radius and the larger (the transverse traceless part of) the anisotropic stress of the source scalar fluctuations around the time of the turn, see \cite{Biagetti:2013kwa,Cai:2019jah} for similar effects in different contexts.
Keeping in mind the bounds on $\etaperp$ coming from ensuring backreaction and perturbative control (sec.~\ref{sec:control}), the upshot is that the contribution to the SGWB from inflationary-generated GWs is subdominant compared to the one studied in the main text, and hence that our results are robust, not being contaminated by the accompanying primordial signal.

\vspace{0.1cm}
\noindent \textbf{Probing primordial features with Stochastic Gravitational Wave Backgrounds.--} Features of the primordial curvature power spectrum are theoretically extremely motivated and detecting them would provide a unique insight into the physics of the early universe. This could for instance pinpoint particular models of inflation, test the inflationary paradigm or point to alternatives to it, or establish the existence of heavy particles beyond the reach of terrestrial experiments \cite{Chen:2010xka,Chluba:2015bqa,Slosar:2019gvt}. To date, observational constraints and prospects for detection have concentrated on the CMB, Large Scale Structure surveys and 21cm tomography (see e.g.~\cite{Chen:2016zuu,Chen:2016vvw,Ballardini:2016hpi,Palma:2017wxu,LHuillier:2017lgm,Ballardini:2017qwq,Beutler:2019ojk,Ballardini:2019tuc} for the latter two). In this context, we see our work as a first step in the direction of probing primordial features with stochastic backgrounds of gravitational waves. Feature models are often broadly classified in three  types: models with sharp features, resonant features, and primordial standard clocks. Our findings here have a broad range of applicability for models falling in the first category, and it would naturally be interesting to study the SGWB signatures of the other types of features.\footnote{See \cite{Braglia:2020taf} for a paper on this subject that appeared after v1 of the present work.} Preliminary investigations of resonant features, mentioned in sec.~\ref{sec:Obs2}, show that they also lead to characteristic features in the SGWB, with GW spectra that exhibit structures periodic in $\log k$. The latter are more complicated than for sharp features, ranging from sinusoidal modulations with frequency $\omegalog$ or $2 \omegalog$, to a double structure of the type~\eqref{eq:resonant_template} in general. 
Scalar-induced SGWBs generated during the radiation era have an energy density power spectrum $h^2 \Omega_\textrm{GW} \sim 10^{-5} {\cal P}^2_{\zeta}$. Hence, the sensitivities of planned GW observatories enable one to probe properties of the curvature fluctuation only if the latter has a typical amplitude on small scales that is boosted compared to CMB and LSS scales. In our paper, this  enhancement is localised in scales and is mainly provided by a copious particle production mechanism. However, one may also envisage features that modulate an otherwise smooth power spectrum of large amplitude on its own (say ${\cal P}_\zeta \sim 10^{-5}-10^{-2}$). Keeping in mind the `averaging-out' effect explained in this paper, it is likely that features in $\Omega_\textrm{GW}$ are appreciable only if the source modulation in the power spectrum is of order one. This by itself can be realised in many different ways. For instance, concerning sharp features, one is not forced to consider copious particle production like studied here, characterised by very large occupation numbers. Instead, more modest particle production with order unity occupation numbers equally generates the required order unity modulation, and hence visible features in $\Omega_\textrm{GW}$.

Independently of the precise setup studied here, we hope our work as well as the general observations above showcase that features in the frequency profile of SGWBs are an interesting theoretically motivated target. Tools have already been developed to successfully reconstruct the spectral shape of mock SGWB signals with LISA \cite{Caprini:2019pxz}, including a wiggly signal. This has been considered as an exotic test case for the reconstruction procedure proposed there, but without particular theoretical motivation. 
The signal on which we have concentrated here has a different frequency profile, and it would be interesting to assess its detectability with LISA and other future GW observatories, as well as more generally to develop dedicated data analysis techniques for probing features with stochastic gravitational wave backgrounds. \\

\acknowledgments
We are grateful to Sebastian Garcia-Saenz, Sadra Jazayeri, Mauro Pieroni and Lucas Pinol for interesting discussions and valuable insights. S.RP would like to thank Gonzalo Palma, Spyros Sypsas and Nicol\'as Parra for useful discussions and collaboration about sharp turns.
J.F, S.RP, and L.T.W are supported by the European Research Council under the European Union's Horizon 2020 research and innovation programme (grant agreement No 758792, project GEODESI).

\appendix

\section{Scalar power spectrum for sharp turns: exact Bogolyubov coefficients, and approximate periodicity} 
\label{sec:analytics}

\subsection{Exact Bogolyubov coefficients}
\label{app:exact-computation}

Here we give the full expressions of the Bogolyubov coefficients, entering into the expression of $\zeta$ \eqref{zeta-out-region} in the \outr\,region, and from which the power spectrum can be computed using Eq.~\eqref{power-spectrum-formal}. Considering the situation in which both types of fluctuations can be considered as massless in the \inr\, region, one finds the following expressions, valid for all scales:
\begin{align}
\label{alpha-zeta} \alpha^\zeta_k &=-\frac{i}{4 \sqrt{1+\X} \etaperp^2} \bigg[\frac{1}{\Sp(\Sp^2-\x^2)} \big(\sum_{\pm}\pm e^{\pm i \deltaeta\, \Sp}\left(1+\etaperp^2(\Sp\mp \x )^2\right) \big)  \\ 
\nonumber & \hphantom{=-\frac{i}{4 \sqrt{1+\X} \etaperp^2} \bigg[}- \frac{1}{\Sm(\Sm^2-\x^2)} \big(\sum_{\pm}\pm e^{\pm i \deltaeta\, \Sm}(1+\etaperp^2(\Sm\mp \x )^2) \big)  \bigg] \, ,\\
\label{beta-zeta} \beta^\zeta_k &=-\frac{ e^{2 i e^{-\delta/2} \x \etaperp}}{2 \sqrt{1+\X} \etaperp^2}
\bigg[ \sum_{\pm} \pm \frac{ \sin( \deltaeta\, \Spm)}{\Spm(\Spm^2-\x^2)}\left(1-2 i \x \etaperp+\etaperp^2(\Spm^2-\x^2)\right)\bigg] \, , \\
\label{alpha-s} \alpha^\s_k&=-\frac{1}{2 \x \etaperp^2}
\bigg[\sum_{\pm} \frac{1}{\Spm(\Spm^2-\x^2)(1+\X\mp \sqrt{\X(1+\X)})} \times \\ 
\nonumber & \hphantom{=-\frac{1}{2 \x \etaperp^2}} \left( (1+2 \x^2 \etaperp^2) \Spm \cos( \deltaeta\, \Spm)+ \etaperp \sin(\deltaeta\, \Spm) \left((1-i \x \etaperp)\Spm^2-(1+i \x \etaperp) \x^2 \right) \right) \bigg], \\
\label{beta-s} \beta^\s_k &=\frac{e^{2 i e^{-\delta/2} \x \etaperp}}{2 \x \etaperp^2}
\bigg[\sum_{\pm} \frac{1}{\Spm(\Spm^2-\x^2)(1+\X\mp \sqrt{\X(1+\X)})} \times \\ 
\nonumber & \hphantom{=\frac{e^{2 i e^{-\delta/2} \x \etaperp}}{2 \x \etaperp^2}
\bigg[} \left( (1-2 i \x \etaperp) \Spm \cos(\deltaeta\,\Spm)+ \etaperp \sin(\deltaeta\, \Spm) (1-i \x \etaperp)(\Spm^2-\x^2) \right) \bigg] \, .
\end{align}
Here, $\x=\frac{k}{\kf \, \etaperp}$ like in the main text, $\X$ is defined in Eq.~\eqref{def-kappa-X}, and $\Spm=\omega_\pm/(H \etaperp)$, where the expressions of the $k$-dependent frequencies $\omega_{\pm}$ are given in Eq.~\eqref{eq:frequencies}. In the main text, for the scales of interest $\x < \sqrt{1-\xi}$ that are exponentially amplified, and for which $\omega_{-}^2<0$, we wrote $S_{-}=i S$, where only $S$ in Eq.~\eqref{def-S} appears in the simplified expressions of interest \eqref{alphas}-\eqref{betas}. There, we used the freedom to factor out an irrelevant $k$-dependent, but \textit{global} phase for all coefficients.

\subsection{Periodicity of peaks for $\xi=-3$}
\label{app:periodicity}

In this appendix we present an iterative method for deriving the period of oscillation in the simplified approximate power spectrum \eqref{eq:P-analytic-xi}. We further specialise to $\xi=-3$, in which case the power spectrum simplifies to \eqref{eq:P-analytic-xi-minus3} which is particularly suited for analytical studies. Here we will focus on the dominant peaks, i.e.~the peaks in the vicinity of the maximum of the envelope of the power spectrum, which for $\xi=-3$ is located at $\x=1$.

Close to $\x=1$ the $\cos$-term in \eqref{eq:P-analytic-xi-minus3} is suppressed compared to the $\sin$-term. Thus as a first approximation the peaks of the power spectrum are given by the maxima of the $\sin$-term. Here we will formalise this and iteratively solve for the positions of the peaks, denoted by $\x_n$, in an expansion about the maxima of the $\sin$-term, denoted by $\x_{0,n}$. This will be an expansion in powers of $(\x_{0,n}-1)$, as we will see explicitly later.

The maxima of the $\sin$-term in \eqref{eq:P-analytic-xi-minus3} are given by 
\begin{align}
\label{eq:maxima-of-sin}
\x_{0,n} = \frac{(2n+1) \pi}{4} \, \frac{e^{\delta /2}}{\etaperp} \, , \quad \textrm{with} \quad n \textrm{ even} \, ,
\end{align}
with odd $n$ corresponding to the minima. We then write $\x_n$ as an expansion about $\x_{0,n}$ as
\begin{align}
\label{eq:xn-expansion}
\x_n = \x_{0,n} + \x_{1,n} = 1 + (\x_{0,n}-1) + \x_{1,n} \, ,
\end{align}
with $\x_{1,n}$ the small correction.
To determine the maxima of $\mathcal{P}(k)$ we need to solve $\mathcal{P}'(k)=0$, which can be brought into the form
\begin{align}
\label{eq:extrema-eq} \cos \big(2 e^{-\delta/2} \etaperp \x \big) + f(\x) \sin \big(2 e^{-\delta/2} \etaperp \x \big) + h(\x) =0 \, ,
\end{align}
where the functions $f(\x)$ and $h(\x)$ can be straightforwardly computed from the expression for $\mathcal{P}(k)$ in \eqref{eq:P-analytic-xi-minus3}.

We can then obtain a solution for $\x_n$ at first order in our expansion by inserting \eqref{eq:xn-expansion} into \eqref{eq:extrema-eq} and solving to first order in $\x_{1,n}$. Here it will be convenient to expand the sinusoidal functions about $\x=\x_{0,n}$, while the functions $f(\x)$ and $h(\x)$ will instead be expanded about $\x=1$, resulting in considerably simpler expressions. The difference between the two expansion loci is $(\x_{0,n}-1)$, which is exactly our expansion parameter, and we will need to keep track of powers of $(\x_{0,n}-1)$ in what follows. Applying this strategy, at first order eq.~\eqref{eq:extrema-eq} is solved by
\begin{align}
\label{eq:maxima-correction-1st}  \x_{1,n} =- \frac{2 e^{-\frac{\delta}{2}} \etaperp + (4\deltaeta -3)}{4 e^{-\frac{\delta}{2}} \etaperp (e^{-\frac{\delta}{2}} \etaperp +1) + (4\deltaeta -3)} (x_{0,n}-1) \, ,
\end{align}
Note that $\x_{1,n} \sim (\x_{0,n}-1)$ which is thus consistent with our assumption of an expansion in powers of $(\x_{0,n}-1)$. The period including the 1st-order correction is thus:
\begin{align}
\label{eq:period-near-max-1st} \Delta \x=\x_{n+2}-\x_{n} &= \pi \frac{e^{\frac{\delta}{2}}}{\etaperp} \bigg(1- \frac{2 e^{-\frac{\delta}{2}} \etaperp + (4 \deltaeta -3)}{4 e^{-\frac{\delta}{2}} \etaperp (e^{-\frac{\delta}{2}} \etaperp +1) + (4 \deltaeta -3)} \bigg) \, ,
\end{align}
which is independent of $n$, i.e.~at 1st-order the modification of the period is a constant shift.  

Using the method laid out here we can go beyond 1st order by further perturbing about the 1st-order corrected solution and solving \eqref{eq:extrema-eq} to higher order in $(\x_{0,n}-1)$. An explicit computation shows that the next correction appears at order $(\x_{0,n}-1)^3$ with the correction depending explicitly on $n$ and thus introducing a drift in the period. However due to the $(\x_{0,n}-1)^3$-suppression this drift only becomes relevant for the peaks away from $\x=1$ whose amplitude is exponentially suppressed compared to the central peaks. Thus, to a high degree of accuracy, the domainant peaks of the power spectrum \eqref{eq:P-analytic-xi-minus3} can be taken as exactly periodic with constant separation \eqref{eq:period-near-max-1st}.

\bibliographystyle{apsrev4-1}
\bibliography{Biblio-2020}
\end{document}